\documentstyle[11pt,psfig,seceq]{article}    

\newcommand{\be}{\begin{equation}}
\newcommand{\ee}{\end{equation}} 
\newcommand{\lb}{\label}

\newcommand{\bF}{{\bf f}}
\newcommand{\bh}{{\bf h}}
\newcommand{\bk}{{\bf k}}
\newcommand{\bp}{{\bf p}}
\newcommand{\bv}{{\bf v}}

\newcommand{\bx}{{\bf x}}
\newcommand{\bom}{{\mbox{\boldmath $\omega$}}}
\newcommand{\bxi}{{\mbox{\boldmath $\xi$}}}
\newcommand{\grad}{{\mbox{\boldmath $\nabla$}}}
\newcommand{\btimes}{{\mbox{\boldmath $\times$}}}
\newcommand{\bdot}{{\mbox{\boldmath $\cdot$}}}
\newcommand{\bdots}{{\mbox{\boldmath $:$}}}
\newcommand{\fL}{\overline{{\bf f}}_\ell}
\newcommand{\vL}{\overline{{\bf v}}_\ell}

\newcommand{\oL}{\overline{{{\mbox{\boldmath $\omega$}}}}_\ell}

\newcommand{\pL}{\overline{p}_\ell}

\newcommand{\vLpm}{\overline{{\bf v}}_\ell^\pm}

\newcommand{\oLpm}{\overline{{{\mbox{\boldmath $\omega$}}}}_\ell^\pm}

\newcommand{\vLmp}{\overline{{\bf v}}_\ell^\mp}

\newcommand{\oLmp}{\overline{{{\mbox{\boldmath $\omega$}}}}_\ell^\mp}

\newcommand{\btau}{{\mbox{\boldmath $\tau$}}_\ell}

\begin{document}
\title{The Joint Cascade of Energy and Helicity \\
in Three-Dimensional Turbulence}
\author{Qiaoning Chen$^{1}$, Shiyi Chen$^{1,2,3}$, and Gregory L. Eyink$^{4}$ \\
$$\,\,$$\\
${}^{1}${\it Department of Mechanical Engineering, The Johns Hopkins University,}\\
{\it Baltimore, MD 21218}\\
${}^{2}${\it Center for Nonlinear Studies and Theoretical Division, Los Alamos National 
Laboratory,}\\
{\it Los Alamos, NM 87545}\\
${}^{3}${\it Peking University, P. R. China}\\
${}^{4}${\it Department of Mathematics, University of Arizona, Tucson, AZ 85721}\\
}
\date{ }
\maketitle
\begin{abstract}
Three-dimensional (3D) turbulence has both energy and helicity as inviscid constants of motion.
In contrast to two-dimensional (2D) turbulence, where a second inviscid invariant---the 
enstrophy---blocks the energy cascade to small scales, in 3D there is a {\it joint cascade} 
of both energy and helicity simultaneously to small scales. It has long been recognized that 
the crucial difference between 2D and 3D is that enstrophy is a nonnegative quantity whereas
the helicity can have either sign. The basic cancellation mechanism which permits a joint cascade 
of energy and helicity is illuminated by means of the {\it helical decomposition} of the velocity 
into positively and negatively polarized waves. This decomposition is employed in the present 
study both theoretically and also in a numerical simulation of homogeneous and isotropic 3D turbulence.
It is shown that the transfer of energy to small scales produces a tremendous growth of helicity 
separately in the $+$ and $-$ helical modes at high wavenumbers, diverging in the limit of infinite 
Reynolds number. However, because of a tendency to restore reflection invariance at small scales, 
the net helicity from both modes remains finite in that limit. Since energy and helicity are not 
separately conserved in the $+$ and $-$ modes, there are four ``flux-like'' quantities for both 
invariants, which correspond to transfer either out of large-scales or into small scales and either 
to $+$ helical or to $-$ helical modes. The helicity fluxes out of large-scales in the separate $+$ 
and $-$ channels are not constant in wavenumber up to the Kolmogorov dissipation wavenumber $k_E$ but 
only up to a smaller wavenumber $k_H$, recently identified by Ditlevsen and Giuliani \cite{DG1,DG2}. 
However, contrary to their argument, the {\it net} helicity flux is shown to be constant all the way 
up to the Kolmogorov wavenumber: there is no shorter inertial-range for helicity cascade than for energy 
cascade. The transfer of energy and helicity between $+$ and $-$ modes, which permits the joint cascade, 
is shown to be due to two distinct physical processes, advection and vortex stretching. 

\end{abstract}

\section{Introduction}

Helicity is the signature of parity-breaking in incompressible fluid flows. For example, 
the most general two-point velocity correlation consistent with homogeneity and isotropy but
{\it not} reflection symmetry is easily seen to have the Fourier transform
\be \langle v_i^*(\bk,t)v_j(\bk,t)\rangle = \left(\delta_{ij}-{{k_ik_j}\over{k^2}}\right)
    {{E(k,t)}\over{4\pi k^2}} - i\varepsilon_{ijl} {{k_l}\over{k^2}}{{H(k,t)}\over{8\pi k^2}}
    \lb{2-point} \ee
where $E(k,t),H(k,t)$ are functions of the wavenumber magnitude $k=|\bk|$ only. $\delta_{ij}=1$ if $i=j$; 0, otherwise. $\varepsilon_{ijl}=1$, if $(i,j,l)$ are cyclic; -1, if $(i,j,l)$ are anticyclic; 0, otherwise. It follows 
immediately from (\ref{2-point}) by integration over $\bk$, both directly and after taking 
a cross-product with $i\bk,$ that
\be \int_0^\infty dk\,\,E(k,t)= {{1}\over{2}}\langle |\bv|^2\rangle \,\,\,\,{\rm and}\,\,\,\,
    \int_0^\infty dk\,\,H(k,t)= \langle \bv \bdot \bom \rangle, \lb{int-spectrum} \ee
where $\bom=\grad\btimes\bv$ is the vorticity field. Thus, $E(k,t)$ is the standard energy 
spectrum, but $H(k,t)$ is the spectrum of the {\it helicity}:
\be       H(t) = \int d\bx \,\,\bv(\bx,t)\bdot\bom(\bx,t).   \lb{helicity} \ee
It is only rather recently that it was realized that this quantity, like energy, is a quadratic 
invariant for the inviscid limit of the fluid equations, the incompressible Euler dynamics
\cite{Mor,Mof}. Nonzero mean values of the helicity are now known to occur naturally in a wide variety 
of geophysical flows. For examples, tornadoes and updraft regions in rotating thunderstorms have
helicity densities approaching $10\,\,m/s^2$ and $0.1\,\,m/s^2$, respectively \cite{LET}. For a review, see \cite{MA}.

Since three-dimensional (3D) incompressible Navier-Stokes dynamics thus possesses two quadratic
inviscid constants of motion, kinetic energy and helicity, it is natural to compare it with the situation 
for two-dimensional (2D) incompressible Navier-Stokes dynamics, which possesses also two quadratic 
invariants, the kinetic energy and the enstrophy $\langle\omega^2\rangle$. As was shown 
by Kraichnan \cite{Kr67}, the presence of the additional enstrophy invariant imposes a strong constraint 
on 2D turbulence, effectively blocking the energy cascade from large scales to small scales. As a result, 
there is in 2D a {\it dual cascade}, with energy cascading to large length-scales and with enstrophy 
cascading to small length-scales. The corresponding situation in 3D was first considered by Brissaud et al. 
\cite{BFLLM}, who considered two possibilities for helicity cascades when the large-scales of the flow 
break reflection symmetry. The first possibility was a {\it pure helicity cascade} to small length-scales, 
in which, similar to the 2D case, there would be only a nonzero helicity flux 
$ \delta $ and no energy flux. 
In that case, Kolmogorov scaling arguments give energy and helicity spectra $E(k)\sim \delta^{2/3}k^{-7/3},
\,\,H(k)\sim \delta^{2/3}k^{-4/3}.$ Note that when these spectra are substituted into (\ref{2-point}), 
the second parity-violating term is the same magnitude as the first term and never becomes negligible. 
The second possibility raised by \cite{BFLLM} was of a {\it joint cascade} in which there would be both 
a non-zero flux of energy $\varepsilon$ and also a non-zero flux of helicity $\delta$ together to small 
length-scales. In that case, an assumption that the transfer rates are determined by $\varepsilon$
alone leads to spectra $E(k)\sim \varepsilon^{2/3}k^{-5/3},\,\,H(k)\sim \delta\varepsilon^{-1/3}k^{-5/3}.$
With these spectra, the second term in (\ref{2-point}) becomes smaller than the first term at high
wavenumber, by a factor $1/k$, and reflection symmetry is asymptotically restored. 

It was argued theoretically by Kraichnan in \cite{Kr73} that the second possibility is more plausible, 
based upon consideration of inviscid equilibria and the fact that a non-vanishing energy spectrum $E(k)$ 
does not imply a non-vanishing helicity spectrum $H(k)$. On the contrary, in 2D, the enstrophy spectrum 
$\Omega(k)=k^2 E(k)$ and flux of energy to high-wavenumber always implies a corresponding enstrophy flux 
larger by the factor $k^2$ at wavenumber $k$. Numerical simulations also support the second possibility.
In \cite{AL} an EDQNM calculation found no evidence of a pure helicity cascade and instead
demonstrated the existence of a joint cascade. Note that the assumption on transfer rates
made by \cite{BFLLM} was built into the closure. A direct demonstration of the joint cascade 
picture was first given by \cite{BO97} in a $128^3$ direct numerical simulation of the Navier-Stokes
equation with a large-scale helical forcing. A high Reynolds number was achieved in that simulation 
with hyperviscosity. We have repeated this simulation with a $512^3$ resolution and with 
ordinary viscosity and verified the result \cite{CCEH}. 

In a recent pair of very interesting papers \cite{DG1,DG2}, Ditlevsen and Giuliani have made 
a number of striking new suggestions concerning the cascade state of energy and helicity.
In particular, they have argued for the existence of the joint cascade, but with a shorter
inertial range for helicity than for energy. They have pointed out the existence of a new 
length-scale $\xi \sim (\nu/\delta)^{3/7}\varepsilon^{2/7}$ which becomes, at high Reynolds 
number, much larger than the Kolmogorov scale $\eta \sim \nu^{3/4}\varepsilon^{-1/4}$. In their
picture, the joint cascade state with constant fluxes $\varepsilon,\,\delta$ and spectra 
$E(k)\sim \varepsilon^{2/3}k^{-5/3},\,\,H(k)\sim \delta\varepsilon^{-1/3}k^{-5/3}$ only holds
for a range of wavenumbers $1/L\ll k\ll 1/\xi$. Thereafter, only the energy flux remains 
constant $\sim \varepsilon$ and the spectrum of energy remains $E(k)\sim \varepsilon^{2/3}k^{-5/3}$
but the inertial cascade of helicity is disrupted by viscous dynamics. In \cite{DG2} it has been 
verified that the proposed scenario holds for a popular shell model of turbulence, the GOY model, 
which has both energy-like and helicity-like inviscid invariants. For 3D Navier-Stokes the authors 
argue for their conclusions making important use of a decomposition of the velocity field into 
helical waves \cite{Waleffe}. In their paper \cite{DG1}, the authors call for a numerical test 
of their proposed scenario in 3D incompressible fluids by a direct numerical simulation.

These works of Ditlevsen and Giuliani are the immediate motivators of the present paper. We believe
that the helical wave decomposition (HWD) employed by them is an important tool in the study of the
joint energy-helicity cascade. It allows the contrast between 2D and 3D turbulence to be made 
most sharply. We follow \cite{DG1,DG2} in making use of the HWD, in even a more general form 
\cite{Moffat,Moses}. However, theoretically our conclusions are at odds with the main conclusions
drawn by Ditlevsen and Giuliani in \cite{DG1,DG2}. In particular, we do {\it not} agree that 
the inertial range for helicity should be any shorter than for energy. We find that their argument
neglects important cancellations that occur for the {\it conserved} quantities, as opposed to their
helical components which are not separately conserved. In fact, the situation is considerably 
richer and more complex for 3D Navier-Stokes than for the simple GOY model. We find that there
{\it are} quantities to which the considerations of \cite{DG1,DG2} apply, but others to which 
they do not. In addition to our new theoretical analysis, we also carry out here the numerical study 
by 3D DNS that was requested in \cite{DG1}. By means of it we have verified all of our important
theoretical conclusions. 

The structure of this paper is as follows. In Section 2 we study the spectral fluxes of energy and helicity
in the helical decomposition. We briefly review in subsection 2.1 the helical decomposition
in Fourier space that we employ. In subsection 2.2 we introduce the helical decomposition of the 
energy and helicity spectra and their balance equations. In subsection 2.3 we relate the transfers
in the separate helical channels to the transfers of the conserved quantities. In subsection 2.4
we study the total transfer integrated over wavenumber and investigate, in particular, the 
signs and magnitudes of transfers between $+$ and $-$ helical modes. One important conclusion established 
there is that transfers will be in a direction so as to restore parity-symmetry at high wavenumbers. 
In subsection 2.5 we strengthen our exact analysis in the previous subsection by a Kolmogorov 
scaling analysis, which leads to a more refined picture. Here we develop our main predictions 
for the spectral fluxes in the joint cascade state. In subsection 2.6 we present the results 
on spectral quantities from a $512^3$ direct numerical simulation with helical forcing. Next, 
in Section 3 we turn to an examination of the joint cascade in physical space, using a filtering
approach to distinguish large-scales and small-scales. In the first subsection 3.1 we present the 
helical decomposition in physical space. In subsection 3.2 we apply this decomposition to dynamics 
of filtered quantities. Here we illuminate the important physical mechanisms contributing 
to transfer between $+$ and $-$ helical modes. These are crucial to give the cancellations
which permit the joint cascade. In subsection 3.3 we consider the relation of filtered quantities
to the spectral ones defined earlier. In subsection 3.4 we apply a Kolmogorov scaling analysis 
to predict the average behavior of the novel quantities which appear in the filtering approach.
In subsection 3.5 we present our numerical results on the filtered quantities from the 3D DNS.
Finally, in Section 4 we summarize our conclusions. 

\section{Spectral Transfer}

\subsection{Helical Wave Decomposition}

In \cite{DG1,DG2} use has been made of an expansion of incompressible velocity fields into
helical waves \cite{Waleffe}. We shall just remind the reader here briefly of this decomposition. 
If the fluid flow is contained in a periodic box, then the velocity field and vorticity field 
may be expanded into circularly-polarized or helical waves $\bh_\pm(\bk)e^{i\bk\bdot\bx}$, where 
$\bh_\pm$ are orthonormal and $i\bk\btimes\bh_\pm = \pm |\bk|\bh_\pm$ \cite{Waleffe}:
\be \bv(\bx,t) = \sum_\bk\sum_{s=\pm} a_s(\bk,t)\bh_s(\bk) e^{i\bk\bdot\bx} \lb{v-Hdecomp} \ee
and 
\be \bom(\bx,t) = \sum_\bk \sum_{s=\pm} s|\bk|a_s(\bk,t)\bh_s(\bk)e^{i\bk\bdot\bx} 
                                                                            \lb{om-Hdecomp} \ee
(Note that \cite{Waleffe} took $\bh_s^*(\bk)\bdot\bh_{s'}(\bk)=2\delta_{ss'}$, but we normalize
to omit the factor $2$.) We then define $\bv^\pm := \sum_\bk a_\pm \bh_\pm e^{i\bk\bdot\bx}$ 
so that $\bv=\bv^+ +\bv^-$. For the vorticity we instead use the convention that $\bom^\pm 
= \sum_\bk |\bk| a_\pm \bh_\pm e^{i\bk\bdot\bx}$ so that $\bom= \bom^+ - \bom^-.$ Thus, 
$\bom^\pm =\pm \grad\btimes\bv^\pm$. 

Because the $+$ and $-$ helical modes belong to distinct eigenspaces of the curl operator,
they are always orthogonal. It follows that the quadratic invariants of energy and helicity may be 
written as $E=E^++E^-$ and $H=H^+-H^-$, where
\be E^\pm(t)={{1}\over{2}}\int d^3\bx\,\,|\bv^\pm(\bx,t)|^2,\,\,\,\,
                     H^\pm(t)=\int d^3\bx \,\,\bom^\pm(\bx,t)\bdot\bv^\pm(\bx,t) \lb{EH-Hdecomp} \ee
By the Parseval equality 
\be E^\pm(t)={{1}\over{2}}\sum_\bk |a^\pm(\bk,t)|^2,\,\,\,\,
                     H^\pm(t)=\sum_\bk |\bk| |a^\pm(\bk,t)|^2.  \lb{per-EH-Hdecomp} \ee
We see that in contrast to the net helicity, which may have either sign, the partial helicities
$H^\pm$ are always nonnegative (with our conventions). However, it is only the combinations 
$E=E^++E^-$ and $H=H^+-H^-$ which are conserved by the inviscid dynamics, and not the separate 
terms. This has important consequences for the spectral dynamics in the $\pm$ helical channels,
which we now consider. 

\subsection{The Helical Decomposition of Spectra and Transfer}

We may define nonnegative spectra for the helical modes via  
\be E^\pm(k,t) = \sum_\bp \,\,{{1}\over{2}}\langle 
                  \hat{v}_i^\pm(\bp) \hat{v}_i(-\bp)\rangle \delta(k-|\bp|) \lb{pm-E-spectrum} \ee 
and 
\be H^\pm(k,t) = \sum_\bp \,\,{{1}\over{2}}\langle 
                  \hat{\omega}_i^\pm(\bp) \hat{v}_i(-\bp)\rangle \delta(k-|\bp|). \lb{pm-H-spectrum} \ee
Note that $E(k,t) = E^+(k,t)+E^-(k,t)$ and $H(k,t)=H^+(k,t)-H^-(k,t)$ give the usual energy and helicity 
spectra. 

It is easy to derive for these modal spectra, the following balance equations 
\be \partial_t E^\pm(k,t) = T_E^\pm(k,t)-2\nu k^2E^\pm(k,t) + F_E^\pm(k,t) \lb{pm-E-balance} \ee
and 
\be \partial_t H^\pm(k,t) = T_H^\pm(k,t)-2\nu k^2H^\pm(k,t) + F_H^\pm(k,t) \lb{pm-H-balance} \ee
Here 
\be F_X^\pm(k,t) = \sum_\bp \langle \hat{\xi}_i^\pm(\bp) \hat{f}_i(-\bp)\rangle \delta(k-|\bp|)
                                                                           \lb{pm-FX-spectra} \ee
with $\bxi=\bv$ for $X=E$ and $\bxi= 2\bom$ for $X=H$. This gives, for $X=E$ and $X=H$, 
the input spectra of energy and helicity, respectively, into the positive and negative 
components of the velocity field, by the force acting at large scales. The $\pm$ transfer functions 
are defined similarly as in \cite{BO97}, by
\be T_X^\pm(k,t) = \sum_\bp N^\pm_X(\bp,t) \delta(k-|\bp|) \lb{pm-X-transfer} \ee
for $X=E,H$, where $N_X^\pm(\bp,t)=-\langle \hat{\xi}_i^\pm(\bp)\widehat{[v_j\partial_jv_i]}(-\bp)\rangle$, 
and $\bxi$ has the same meaning as above. This formula will be used for numerical computation 
of $T_X^\pm(k,t)$. It is also a simple consequence of (\ref{pm-FX-spectra}) and (\ref{pm-X-transfer})
that the standard input and transfer spectra are obtained as 
\be F_E(k,t)=F_E^+(k,t)+F_E^+(k,t),\,\,\,\,T_E(k,t)=T_E^+(k,t)+T_E^+(k,t), \lb{E-total-FandT} \ee
and 
\be F_H(k,t)=F_H^+(k,t)-F_H^+(k,t),\,\,\,\,T_H(k,t)=T_H^+(k,t)-T_H^+(k,t). \lb{H-total-FandT} \ee

An alternative set of formulas that are theoretically useful follow from 
\be E^\pm(k,t) = \sum_\bp \,\,{{1}\over{2}}\langle |a^\pm(\bp,t)|^2 \rangle \delta(k-|\bp|) 
                                                               \lb{pm-E-spectrum-a} \ee 
and 
\be H^\pm(k,t) = \sum_\bp \,\, |\bp| \langle |a^\pm(\bp,t)|^2 \rangle \delta(k-|\bp|) 
                                                               \lb{pm-H-spectrum-a}. \ee 
where $a^\pm$ are ($\sqrt{2}$ times) the helical amplitudes defined by Waleffe. See \cite{Waleffe},
Eq.(5). From these we derive the very important relation that 
\be      H^\pm(k,t)= 2k E^\pm(k,t). \lb{pm-maximal} \ee
The well-known inequality for the net spectra that $H(k,t)\leq 2k E(k,t)$ \cite{BFLLM,Kr73} follows 
immediately. We see that $\pm$ modes are separately maximal helicity modes. Eq.(\ref{pm-maximal}) 
is the closest 3D analogue of the relation $\Omega(k,t)=k^2E(k,t)$ which in 2D implies blocking 
of forward energy cascade by enstrophy conservation \cite{Kr67}. 

\subsection{Relation to Transfers of the Inviscid Invariants}

The $\pm$ parts of (\ref{pm-E-balance}),(\ref{pm-H-balance}) may be combined using (\ref{E-total-FandT}),
(\ref{H-total-FandT}) to yield the standard spectral balance equations for energy and helicity:
\be \partial_t E(k,t) = T_E(k,t) -2\nu k^2 E(k,t) + F_E(k,t) \lb{E-balance} \ee
and 
\be \partial_t H(k,t) = T_H(k,t) -2\nu k^2 H(k,t) + F_E(k,t) \lb{H-balance} \ee
However, there is a crucial difference between the transfer functions of the 
conserved quantities, $T_H,T_E,$ on the one hand, and the transfer functions
$T_E^\pm,T_H^\pm,$ on the other hand. The integral relations
\be  \int_0^\infty dk\,\,T_X(k,t) = 0  \lb{X-conserve} \ee
for $X=E,H$ are the statement of conservation of energy and of helicity, respectively.
However, such relations are {\it not} satisfied by $T_E^\pm,T_H^\pm$ separately.

The important relation (\ref{pm-maximal}) allows us to obtain both the transfer functions $T_E^\pm$ 
and $T_H^\pm$ in terms of $T_E,T_H$. Indeed, as previously noted, 
\be     E^+(k,t) + E^-(k,t) = E(k,t)  \lb{E-spec-total} \ee
and from $H^+(k,t)-H^-(k,t)=H(k,t)$ using (\ref{pm-maximal}) 
\be E^+(k,t) - E^-(k,t) = H(k,t)/(2k). \lb{H-spec-total} \ee  
These can be solved to give
\be E^\pm(k,t) = {{1}\over{2}}\left(E(k,t)\pm {{H(k,t)}\over{2k}}\right), \lb{pm-E-spec} \ee
and thus by (\ref{pm-maximal}) again 
\be H^\pm(k,t) = {{1}\over{2}}\left(2k E(k,t)\pm H(k,t)\right). \lb{pm-H-spec} \ee
These equations are well-known \cite{Lesieur}. However, if we then take the time derivative and use 
the balance relations for energy, we find moreover that  
\be T_E^\pm(k,t) = {{1}\over{2}}\left[T_E(k,t)\pm {{T_H(k,t)}\over{2k}}\right], \lb{pm-E-transfer} \ee
and that 
\be T_H^\pm(k,t) = {{1}\over{2}}\left[2k T_E(k,t)\pm T_H(k,t) \right], \lb{pm-H-transfer} \ee
in terms of $T_E,T_H$. Of course, similar relations can also be developed for the forcing spectra 
$F_E^\pm,F_H^\pm$.

We may define altogether four kinds of ``flux-like'' quantities 
\be \Pi_X^{\pm,<}(k) = -\int_0^k dp\,\,T_X^\pm(p) \lb{pm-lt-X-flux} \ee
and 
\be \Pi_X^{\pm,>}(k) = \int_k^\infty dp\,\,T_K^\pm(p) \lb{pm-gt-X-flux} \ee
for each $X=E,H$. Thus, $\Pi_X^{\pm,<}(k)$ represents the flow of $X$ out of $\pm$ modes for the wavenumbers 
$<k$ and $\Pi_X^{\pm,>}(k)$ represents the flow of $X$ into the $\pm$ modes for the wavenumbers $>k$. 
These quantities are generally {\it not} equal for $<$ and $>$. In contrast,
the net fluxes $\Pi_E^{\Join}= \Pi_E^{+,\Join}+\Pi_E^{-,\Join},\,\,\Pi_H^{\Join}= 
\Pi_H^{+,\Join}-\Pi_H^{-,\Join}$ for $\Join=<$ or $>$, satisfy  
\be \Pi_X^{<}(k) = \Pi_X^{>}(k) \lb{lt-gt-same} \ee
for $X=E,H$, as a consequence of (\ref{X-conserve}). Therefore, no distinction need be made between 
the $<$ and $>$ quantities for the net fluxes. However, for the $+$ and $-$ modes separately the $<$ and $>$
fluxes may be different. We shall see that this is indeed the case. 

\subsection{Total Balances in the Helical Decomposition}

A number of important qualitative results may be derived from the expressions (\ref{pm-E-transfer}),
(\ref{pm-H-transfer}) for the total balances in the $\pm$ modes over all wavenumbers. If we integrate the 
relations $(\ref{pm-E-balance})$ and $(\ref{pm-H-balance})$ over wavenumber, we obtain 
\be {{dE^\pm}\over{dt}}= \pm R^E - D_E^\pm + F_E^\pm   \lb{tot-pm-E-balance} \ee
and
\be {{dH^\pm}\over{dt}}= R^H - D_H^\pm + F_H^\pm  \lb{tot-pm-H-balance} \ee
where $D_E^\pm(t):= 2\nu \int_0^\infty dk\,\,k^2E^\pm(k,t)$ and $D_H^\pm(t):=
2\nu \int_0^\infty dk\,\,k^2H^\pm(k,t)$ are the total dissipation of energy 
and helicity, respectively, in the $+$ and $-$ components, and where 
$F_E^\pm(t) = \int_0^\infty dk\,\, F_E^\pm(k,t)$ and $F_H^\pm(t) = \int_0^\infty dk
\,\, F_H^\pm(k,t)$ are the total forcing inputs of energy and helicity, respectively, 
in the $+$ and $-$ components. For $R_E,R_H$ we obtain the expressions 
\be R_E:= \pm \int_0^\infty dk\,\,T_E^\pm (k,t)= \int_0^\infty dk\,\,
                           {{T_H(k,t)}\over{4k}},    \lb{tot-E-transfer} \ee
and 
\be R_H:= \int_0^\infty dk\,\,T_H^\pm (k,t)=  \int_0^\infty dk\,\, kT_E(k,t) \lb{tot-H-transfer} \ee
by integrating (\ref{pm-E-transfer}),(\ref{pm-H-transfer}) over wavenumber and using (\ref{X-conserve}). 

As (\ref{tot-pm-E-balance}),(\ref{tot-pm-H-balance}) show, the $+$ and $-$ components of energy and 
helicity are not separately conserved (in the absence of forcing and dissipation). In general, there 
is no reason that $R_E=0$ or $R_H=0$. Instead, there will be a nonzero transfer between $+$ and $-$ 
components, $R_E$ for energy and $R_H$ for helicity, which exactly balance, so that total energy and 
helicity are conserved. This has an interesting consequence. For total energy and helicity, 
the forcing input and the dissipation must balance, in the steady state:
\be D_E = F_E=\varepsilon,\,\,\,\,\,\,\,\, D_H=F_H=\delta. \lb{tot-D-F-balance} \ee
However, for the $+$ and $-$ components separately, the input and dissipation
do {\it not} balance, and in general there is a discrepancy:
\be D_E^\pm-F_E^\pm = \pm R_E,  \,\,\,\,\,\,\,\,D_H^\pm-F_H^\pm = R_H. \lb{pm-EH-discrep} \ee

The signs of $R_E,R_H$ can be inferred from (\ref{tot-E-transfer}),(\ref{tot-H-transfer}). If the 
large-scales contain mostly positive helicity, then $T_H(k)<0$ at low wavenumbers and (\ref{tot-E-transfer})
implies $R_E<0$; conversely, if the large-scales contain mostly negative helicity, then $T_H(k)>0$ 
at low wavenumbers and (\ref{tot-E-transfer}) implies that $R_E>0.$ Therefore, if we input for 
example mostly positive helicity at large scales, then $D_E^+<F_E^+$ and $D_E^->F_E^-$. This means 
that some of the energy carried by the $+$ modes at large scales will be converted to energy of the $-$ 
modes during transfer, and will then be accounted by the dissipation $D_E^-$ rather than by $D_E^+$. 
We see that quite general considerations imply that there is a tendency toward equalization 
of energy in the $+$ and $-$ components due to the nonlinear transfer between them. Hence,
reflection-symmetry which is broken at large scales will tend to be restored asymptotically at small scales.

A similar argument using the formula (\ref{tot-H-transfer}) for $R_H$ implies that $R_H>0$ always, 
because $T_E(k)>0$ at high wavenumbers in any case. Therefore, viscous dissipation of helicity 
in $+$ and $-$ components in general {\it both} exceed their forcing inputs: $D_H^\pm > F_H^\pm$. 
As the energy cascades to high wavenumber in each compoment, a corresponding helicity is carried 
in each component $\sim 2k E^\pm(k)$ at wavenumber $k$, which in fact grows to exceed the input helicity 
in that component at large enough $k$. This is the same mechanism that, in 2D turbulence, prevents 
energy cascade to high wavenumbers due to enstrophy conservation. However, helicity has both signs, 
so that in 3D the growth in $H^+(k)$ is compensated by an equal growth in $H^-(k)$
at high wavenumbers, with the net helicity conserved. The consequence is that {\it both} $D_H^+$ 
and $D_H^-$ will exceed their inputs. The equal growth of helicity in $+$ and $-$ modes at
high wavenumber again implies an asymptotic restoration of parity-symmetry.

\subsection{Kolmogorov Scaling Theory}

A clearer picture is obtained from a Kolmogorov scaling analysis for steady-state helical 
turbulence in 3D. A similar analysis was attempted by Ditlevsen and Giuliani in \cite{DG1}
but our assumptions and conclusions differ in some important respects from theirs. Ignoring 
for the moment any intermittency corrections, we assume that 
\be E(k) \sim C_E \varepsilon^{2/3} k^{-5/3}  \lb{K41-E-spectrum} \ee
and 
\be H(k) \sim C_H (\delta/\epsilon^{1/3}) k^{-5/3} \lb{K41-H-spectrum} \ee
both for $k_L\ll k \ll k_E \sim \varepsilon^{1/4}\nu^{-3/4}$. $k_E$ is the standard Kolmogorov 
wavenumber. The joint $5/3$ spectra for energy and helicity were first proposed in \cite{BFLLM,Kr73}. 
The main assumption behind (\ref{K41-H-spectrum}) is that the transfer time for helicity at wavenumber 
$k$ is the same as for energy, $\tau_H(k)\sim \tau_E(k) \sim \varepsilon^{-1/3} k^{-2/3}$ \cite{BFLLM}. 
Equations (\ref{K41-E-spectrum}),(\ref{K41-H-spectrum}) along with (\ref{pm-E-spec}) imply the 
corresponding energy spectra for components, already noted in \cite{DG1,Lesieur}: 
\be E^\pm(k) \sim {{1}\over{2}}C_E \varepsilon^{2/3} k^{-5/3} \left[ 1\pm {{\gamma}\over{2k}} 
                 \left({{\delta}\over{\varepsilon}}\right)\right], \lb{K41-pm-E-spectra} \ee
with $\gamma=C_H/C_E$, and from (\ref{pm-H-spec}) the helicity spectra
\be H^\pm(k) \sim C_E \varepsilon^{2/3} k^{-2/3} \left[ 1\pm {{\gamma}\over{2k}} 
                 \left({{\delta}\over{\varepsilon}}\right)\right], \lb{K41-pm-H-spectra} \ee
for $k_L\ll k \ll k_E.$ However, Ditlevsen and Giuliani in \cite{DG1} arrive at a conclusion 
in contradiction to our starting assumptions (\ref{K41-E-spectrum}),(\ref{K41-H-spectrum}). 
They define an intermediate wavenumber $k_H\sim \varepsilon^{-2/7}(\delta/\nu)^{3/7}$ 
which is $\ll k_E$ at high Reynolds number. We refer to $k_H$ hereafter as the ``Ditlevsen wavenumber''.
It is concluded in \cite{DG1} that only for ``the inertial range $K[={\rm our}\,\,k_L]<k<k_H$ there 
is a coexisting cascade of energy and helicity where helicity follows a `linear cascade' with a 
$H(k)\sim k^{-5/3}$ spectrum. In the range $k_H<k<k_E$ the dissipation of helicity dominates with 
a detailed balance between dissipation of positive and negative helicities and the right-left symmetry 
of the flow is restored.'' We agree essentially with the latter statement, but we argue that the joint 
cascade of energy and helicity and the spectrum (\ref{K41-H-spectrum}) hold up to the 
Kolmogorov wavenumber $k_E$, not up just to the Ditlevsen wavenumber $k_H$. 

We demonstrate the consistency of our assumptions. According to (\ref{K41-pm-E-spectra}), 
the total energy separately in the $\pm$ modes $\int_0^{k_E} dk\,\,E^\pm(k)$ both remain 
finite as $\nu\rightarrow 0$, but by (\ref{K41-pm-H-spectra}) the helicity in the separate 
$\pm$ modes increase as 
\be \int_0^{k_E} dk\,\,H^\pm(k) \sim \varepsilon^{3/4}\nu^{-1/4} \lb{pm-H-diverge} \ee
for small $\nu$. Thus, both of these latter integrals will diverge in the limit of
infinite Reynolds number. However, the total helicity $\int_0^{k_E} dk\,\,H(k)$
will remain finite. This agrees with our conclusions in the previous subsection that helicity
in the $\pm$ modes separately will be produced as a consequence of energy transfer to 
higher wavenumbers in those modes. We may also consider total integrated dissipations. 
We see that in each of the $\pm$ modes separately 
\be D_E^\pm = 2\nu\int_0^{k_E} dk\,\,k^2 E^\pm(k) \sim \nu \varepsilon^{2/3}
             k_E^{4/3}\propto \varepsilon, \lb{K41-pm-E-dissipation} \ee 
whereas
\be D_H^\pm = 2\nu\int_0^{k_E} dk\,\,k^2 H^\pm(k) \sim \nu \varepsilon^{2/3}
             k_E^{7/3}\propto \varepsilon^{5/4}\nu^{-3/4}, \lb{K41-pm-H-dissipation} \ee 
diverging as $\nu\rightarrow 0$. This conclusion was already reached in \cite{DG1,DG2}.
However, we see with our assumption (\ref{K41-H-spectrum}) that 
\be D_H  = 2\nu\int_0^{k_E} dk\,\,k^2 H(k) \sim \nu (\delta/\varepsilon^{1/3})
             k_E^{4/3}\propto \delta, \lb{K41-H-dissipation} \ee 
balancing the helicity input from the force, as it must. The important contribution to this 
integral comes from $k\sim k_E$. On the contrary, Ditlevsen and Giuliani in \cite{DG2}
argued that ``the integral will not be dominated by contributions from $k_E$ but contributions
from $k_H$.'' This statement ignores precisely the cancellation which occurs between $+$ and 
$-$ modes in (\ref{K41-H-dissipation}). 

However, the Ditlevsen wavenumber does play an important role in the behavior of the flux 
quantities $\Pi_H^{\pm,<}(k)$ and $\Pi_H^{\pm,>}(k)$ defined earlier. We show that {\it both} 
of these fluxes have constant values over a range of wavenumbers $k_L\ll k \ll k_H$ (at least),
but that the constants are different. In fact, if we assume that the forcing is confined to a finite 
range of low wavenumbers, then integrating the relation (\ref{pm-H-balance}) over the wavenumber 
interval $[0,k]$ yields
\be F_H^\pm = \Pi_H^{\pm,<}(k) + 2\nu \int_0^k dp\,\,p^2 H^\pm(p). \lb{pm-lt-H-flux} \ee
Similarly, integrating (\ref{pm-H-balance}) over the wavenumber interval $[k,\infty]$ yields
\be D_H^\pm = \Pi_H^{\pm,>}(k) + 2\nu \int_0^k dp\,\,p^2 H^\pm(p). \lb{pm-gt-H-flux} \ee
Now $2\nu\int_0^{k} dp\,\,p^2 H^\pm(p) \sim \nu \varepsilon^{2/3} k^{7/3}$ so
that this integral $\propto \delta$ precisely when $k\sim k_H$ as defined above. 
Since $F_H^\pm$ are generally of the same order of magnitude as $\delta=F_H^+-F_H^-,$ 
\be \Pi_H^{\pm,<}(k) \approx F_H^\pm,  \lb{pm-lt-H-flux-const} \ee
for the range $k_L\ll k \ll k_H$, smaller than the conventional inertial range 
$k_L\ll k \ll k_E$. On the other hand, 
\be \Pi_H^{\pm,>}(k) \approx D_H^\pm  \lb{pm-gt-H-flux-const} \ee
over the conventional range, because the error term only becomes comparable to
$D_H^\pm$ when $k$ approaches $k_E$. As we have seen, $D_H^\pm > F_H^\pm$ in general, 
with $D_H^\pm$ becoming arbitrarily large as Reynolds number increases. For $k>k_H$ 
and $k>k_E$, respectively, both the $<$ and $>$ flux-like quantities decrease below 
their ``inertial-range'' values, $F_H^\pm$ and $D_H^\pm$. In fact, $\Pi_H^{\pm,>}(k)$ 
decreases from $D_H^\pm$ to $0$, while $\Pi_H^{\pm,<}(k)$ decreases from $F_H^\pm$ 
to the {\it negative} value $F_H^\pm-D_H^\pm =-R_H$. 

Ditlevsen and Guliani argued in \cite{DG2} that {\it net} helicity flux 
\be \Pi_H(k) = \Pi_H^{+,<}(k)-\Pi_H^{-,<}(k) = \Pi_H^{+,>}(k) -\Pi_H^{-,>}(k) \lb{net-H-flux} \ee
would also have a constant value in only the same range, $k_L\ll k \ll k_H$, 
which is smaller than the conventional inertial range, $k_L\ll k \ll k_E$. They illustrated 
this phenomemon in \cite{DG2} for the GOY shell model. However, their argument ignores the cancellation 
that takes place between $+$ and $-$ components for Navier-Stokes. In fact, for net helicity flux, 
\be F_H  = \Pi_H(k) + 2\nu \int_0^k dp\,\,p^2 H(p) \lb{lt-H-flux} \ee
and 
\be D_H = \Pi_H(k) + 2\nu \int_0^k dp\,\,p^2 H(p). \lb{lt-E-flux} \ee
and the latter integral, $2\nu \int_0^k dp\,\,p^2 H(p)$, is negligible for all 
wavenumbers $k$ in the range $k_L\ll k \ll k_E$. Thus, $F_H\approx \Pi_H(k)
\approx D_H$ for all $k$ in the conventional inertial range, as expected. 

Exactly the same arguments as made above for helicity, when applied 
to energy, show that 
\be F_E^\pm = \Pi_E^{\pm,<}(k) + 2\nu \int_0^k dp\,\,p^2 E^\pm(p) \lb{pm-lt-E-flux} \ee
and 
\be D_E^\pm = \Pi_E^{\pm,>}(k) + 2\nu \int_0^k dp\,\,p^2 E^\pm(p). \lb{pm-gt-E-flux} \ee
Thus, for the range $k_L\ll k \ll k_E$,
\be \Pi_E^{\pm,<}(k) \approx F_E^\pm, \lb{pm-lt-E-flux-const} \ee
but
\be \Pi_E^{\pm,>}(k) \approx D_E^\pm. \lb{pm-gt-E-flux-const} \ee
The situation is quite similar to that for helicity, except constancy of 
these flux-like quantities holds over the whole extent of the conventional 
inertial range and the discrepancy $D_E^\pm-F_E^\pm$ does not grow with 
the Reynolds number, but tends to a finite limit.

\subsection{Simulation Results}

To order to test the various theoretical predictions in subsection 2.1 $\sim$ 2.5, we have carried out direct numerical simulation of incompressible 
isotropic turbulence by solving the forced Navier-Stokes equations using a pseudospectral calculation with $512^3$ resolution. The system was forced 
by holding the kinetic energy fixed in the first two shells for the wavenumber $|\bk|<2.5$ \cite{Shiyi}. 
Nonnegative helicity was added by using a representation of the helicity spectrum in \cite{PolShtil}:
\be H(k,t) = \sum_{\bk'\in S(k)} 2\bk'\bdot [{\bf R}(\bk',t)\btimes{\bf I}(\bk',t)] \lb{Pol-Sht} \ee
where $S(k)$ is the wavenumber shell $\{\bk': k-(\Delta k/2)<|\bk'|<k+(\Delta k/2)\}$ and ${\bf R}(\bk,t)$
and ${\bf I}(\bk,t)$ denote the real and imaginary parts of the Fourier-transformed velocity
$\widehat{\bv}(\bk,t)$, respectively. At each time-step we rotated ${\bf I}(\bk,t)$ for all modes
$|\bk|<2.5$ so that its angle with ${\bf R}(\bk,t)$ was kept at $90^\circ$ with positive handedness. 
This method was used by us earlier in \cite{CCEH}. It guarantees a positive helicity spectrum
in the low wavenumbers and  $H(k,t)<2kE(k,t)$ in general. The microscale Reynolds number 
achieved in our simulation was $R_\lambda=220.$ The data analyzed were obtained after the system reached 
statistical equilibrium. The values of inputs of energy and helicity were $\varepsilon=0.1427$ and 
$\delta=0.4415$. 

In Fig.~1 are plotted the spectral fluxes of energy and helicity.

\noindent Clearly, there is about a decade of inertial range where these fluxes have a constant mean value.
It can be seen also that the intervals of constant flux for energy and helicity have about the same 
range in Fourier space, contrary to the prediction of \cite{DG1} but in agreement with our modified 
theory. To really test this, we should increase the Reynolds number and show that the constant flux 
ranges for both invariants grow with the Reynolds number in the same manner. Computational limitations
unfortunately prevent us from raising the Reynolds number beyond the present value. However, further 
evidence that the two constant flux ranges have the same extent will be given below for a shell model, 
where our arguments about cancellation also apply. 

Let us next consider the behavior in the separate $\pm$ helical modes. Some of the 
relevant parameter values in our simulations are given in Table 1. 
The values of the dissipations $D_E^\pm,D_H^\pm$ were obtained from the integrals 
(\ref{K41-pm-E-dissipation}) and (\ref{K41-pm-H-dissipation}), respectively. The transfers $R_E,R_H$ we calculated from the integrals 
(\ref{tot-E-transfer}) and (\ref{tot-H-transfer}), respectively. With our forcing technique, the inputs $F_E^\pm,F_H^\pm$
cannot be similarly obtained, because the forcing spectra are not easily computable. Instead, 
we have obtained those values via the relations (\ref{pm-EH-discrep}), i.e. $F_E^\pm =D_E^+\mp R_E$ 
and $F_H^\pm=D_H^\pm-R_H$. 
We wish to examine the predictions made in section 2.5 for the fluxes $\Pi_X^{\pm,<},\Pi_X^{\pm,>}$ 
defined in (\ref{pm-lt-X-flux}),(\ref{pm-gt-X-flux}) for $\pm$ polarization and $X=E,H$. We have 
predicted both the extents of the inertial ranges for these quantities and the plateau values 
in these ranges.

\begin{center}
\begin{table}
\caption{Parameters of the Flow}
\begin{center}
\begin{tabular}{||c|c||}
\hline
Energy    & Helicity  \\ 
\hline
$D_E^+ = 0.0723$ & $D_H^+=7.30$ \\
$D_E^- = 0.0685$ & $D_H^-=6.86$ \\
$R_E = -0.05152$ & $R_H=6.67$ \\
$F_E^+=0.1239$ & $F_H^+=0.63$ \\
$F_E^-=0.01698$ & $F_H^-=0.19$ \\
$\varepsilon=0.1427$ & $\delta=0.4415$ \\    
\hline
\end{tabular}
\label{tb.k1}
\end{center}
\end{table}
\end{center}

In Fig.~2 (a) and (b) we plot $\Pi_E^{\pm,>}$ and  $\Pi_E^{\pm,<}$. 

\noindent We have predicted that {\it all} of these fluxes 
shall be constant over the same range $k_L\ll k \ll k_E$ in the limit of high Reynolds number. 
It seems from the figure, however, that $\Pi_E^{-,<}$ has a somewhat shorter range than the 
other three fluxes. This does not contradict our predictions, since those were asymptotic 
statements at high Reynolds number. More precisely, the prediction is that all of the fluxes shall 
have ranges which grow at the same rate in that limit. This does not rule out their having 
ranges with lengths that differ by a constant factor. In fact, it is easy to see why $\Pi_E^{-,<}$ 
appears to have a shorter range in Figure 2. Note from the table above that $F_E^-$ is an order 
of magnitude smaller than $F_E^+$ in our simulation. The dissipation term in (\ref{pm-lt-E-flux}) 
is negligible compared to the input value $F_E^\pm$ for wavenumbers $k\ll k_E^\pm \sim 
(F_E^\pm/\varepsilon^{2/3}\nu)^{3/4}$, respectively, for $\pm$ helical modes. More precisely,
the wavenumber $k_E^\pm$ can be defined as that for which the dissipation integral reaches a fixed
fraction, say, $99\%$, of the input $F_E^\pm$. Although $k_E^\pm$ grow at the same rate with Reynolds 
number, we see that, with our parameter values, $k_E^-$ should be only about 1/5 as big as $k_E^+$. 
This agrees very well with the results in Figure 2. 

Next let us consider scaling ranges for helicity. In Fig.~3 (a)and (b) we plot $\Pi_H^{\pm,>}$ and 
$\Pi_H^{\pm,<}$. 

\noindent We have predicted that $\Pi_H^{\pm,>}$ are constant from the injection wavenumber $k_L$ up to the 
Kolmogorov wavenumber $k_E$, while $\Pi_H^{\pm,<}$ are constant only up to the Ditlevsen wavenumber 
$k_H$ (within constant factors). We can indeed see that the ranges for $\Pi_H^{\pm,<}$ are shorter. 
More quantitatively, note that $k_H/k_E\sim (k_L/k_E)^{3/7}$ and since $k_L/k_E\sim 0.1$ in our simulation, 
we calculate that $k_H/k_E$ is about $1/3$. This is quite consistent with the ranges observed in Figure 3. 
In contrast to the energy, {\it both} $\Pi_H^{+,<}$ {\it and} $\Pi_H^{-,<}$ have ranges shorter than the 
conventional inertial range, and they are expected to grow at a different rate than the Kolmogorov 
wavenumber. Furthermore, close examination of Figure 3 shows that the range for $\Pi_H^{-,<}$ seems even 
a little shorter still than the range for $\Pi_H^{+,<}$. Arguing just as for the energy, we see that 
the dissipation term in (\ref{pm-lt-H-flux}) is negligible compared with $F_H^\pm$ for wavenumbers 
$k\ll k_H^\pm \sim (F_H^\pm/\varepsilon^{2/3}\nu)^{3/7}$, respectively, for $\pm$ helical modes. Because 
$F_H^-$ is about 3 times smaller than $F_H^+$ in our simulation, we predict that $k_H^-$ should be only 
about $2/3$ as big as $k_H^+$. This is consistent with Figure 3. 

To verify our prediction for the plateau values of the fluxes, in Fig.~4(a) and (b) we plot $\Pi_E^{\pm,>}$ and  $\Pi_E^{\pm,<}$. 

\noindent In Fig.~4 (a), we can see 
that $\Pi_E^{\pm,>}$ drop from values very close to $D_E^{\pm}$ in the inertial range to zero in the 
high wavenumber range, while we see in Fig.~4 (b) that $\Pi_E^{\pm,<}$ decrease from values 
close to the forcing inputs $F_E^{\pm}$ in the inertial range to values $\mp R_E$ in the small scales. 
Likewise, in Fig.~5(a)(b) we plot $\Pi_H^{\pm,>}$ and  $\Pi_H^{\pm,<}$. 

\noindent There $\Pi_H^{\pm,>}$ drops from a value close to $D_H^{\pm}$ in the inertial range to zero at high 
wavenumber, and $\Pi_H^{\pm,<}$ from constant values at low wavenumbers to $-R_H$ at high wavenumbers. Notice 
that these constants are somewhat smaller than the forcing inputs $F_H^{\pm}$, especially in $-$ modes, 
which is different from the prediction in (\ref{pm-lt-H-flux-const}). The reason is that the inertial 
range in our simulation is not long enough so that we can not neglect the dissipation term in 
(\ref{pm-lt-H-flux}). By calculating the dissipation explicitly, we have verified that the observations 
are consistent with (\ref{pm-lt-H-flux}).

It is clear that all of our theoretical predictions are consistent with the reported numerical
simulation results, both qualitatively and quantitatively. Figure 1 
demostrates the prediction that energy and helicity fluxes are constant over ranges 
of the same extent. Our results contradict the predictions of Ditlevsen and Giuliani that helicity flux shall
be constant only up to the wavenumber $k_H\sim \varepsilon^{-2/7}(\delta/\nu)^{3/7}$ \cite{DG1}. 
Since only a short inertial range can be achieved in our 3D simulations, it is worthwhile to give 
further evidence in a model with a longer inertial range. We therefore consider a {\it helical shell model}, 
which, like 3D Navier-Stokes has both $+$ and $-$ helical modes and energy and helicity as inviscid invariants 
\cite{BifKerr,BBKT}. We consider a SABRA version of this model \cite{SABRA}, because the scaling properties are 
a little better than in the original GOY-type models. The model we study is, precisely,  
\begin{eqnarray} 
\dot{u}_n^+ & = & i\left[ a k_n u_{n+2}^- (u_{n+1}^+)^* + b k_{n-1} u_{n+1}^- (u_{n-1}^+)^* \right. \cr
         \, &   &  \left. - c k_{n-2}  u_{n-1}^- u_{n-2}^- \right] - \nu k_n^2 u_n^+ + f_n^+ \lb{SABRA3} 
\end{eqnarray}
\begin{eqnarray}
\dot{u}_n^- & = & i\left[ a k_n u_{n+2}^+ (u_{n+1}^-)^* + b k_{n-1} u_{n+1}^+ (u_{n-1}^-)^* \right. \cr
         \, &   &  \left. - c k_{n-2}  u_{n-1}^+ u_{n-2}^+ \right] - \nu k_n^2 u_n^+ + f_n^- \lb{2SABRA3}
\end{eqnarray} 

\noindent for a set of modes $u_n^+$ and $u_n^-,\,\,n=1,...,N$. Here the wavenumber $k_n = \lambda^n k_0$ for some $\lambda>1$. This model may be 
termed SABRA3 in the classification scheme analogous to \cite{BBKT}. If $a+b+c=0$, then this model for 
$\nu=f=0$ conserves both an ``energy'' $E$ and ``helicity'' $H$, where
\be E(t)=\sum_{n=1}^N {{1}\over{2}}[|u_n^+|^2 + |u_n^-|^2], \,\,\,\, 
                                  H(t)=\sum_{n=1}^N k_n^\alpha [|u_n^+|^2 -|u_n^-|^2] \lb{shell-EH} \ee
and $x=\lambda^\alpha$ is the second root of the quadratic $ax^2+bx+c=0$, in addition to $x=1$. We consider 
a standard choice of parameters $a,b,c$ given in Table II of \cite{BBKT}, with $\lambda=2$ and $\alpha=1$. 
Thus, the second invariant is truly ``helicity-like''. 

Unlike the GOY shell model considered by \cite{DG1}, the SABRA3 model considered here can show the same 
cancellations between $+$ and $-$ modes that we have argued to occur in Navier-Sokes. If the argument of 
\cite{DG1} for Navier-Stokes were correct, it would apply also to this helical shell model. To check their 
prediction, we could vary either $\nu$ or $\delta$ and observe the changes in the lengths of the constant flux ranges. 
In fact, it is easiest to change $\delta$. We performed two 22-shell simulations for SABRA3 with $\nu=10^{-7}$, 
one with helicity input $\delta=0.0015$ and another with helicity input $\delta=0.0003$, 5 times smaller. 
If the prediction of \cite{DG1} were correct, then going from the first to the second, the inertial range 
for energy flux should be unchanged but the inertial range for helicity flux should be 2 times shorter. 
Fig.~6 shows the helicity flux in the two simulations with different helicity inputs. 

\noindent In the inset are the normalized energy and helicity fluxes. As we can see very
convincingly from this plot, {\it none} of the constant flux ranges has its length changed 
by changing helicity input $\delta$. This is in agreement with our prediction, but in disagreement 
with the prediction of \cite{DG1}.

\newpage

\section{Filtering Approach}

\subsection{Helical Decomposition in Physical Space}

In \cite{DG1,DG2} use has been made of an expansion of incompressible velocity fields into
helical waves, for flow in a periodic box or infinite space \cite{Waleffe}. In fact, the helical 
decomposition is an intrinsic decomposition and can be made for any flow domain $\Lambda$. It corresponds
just to a decomposition of the velocity field $\bv(\bx,t)$ into eigenmodes of the non-dimensionalized 
curl operator $\Sigma=(-\bigtriangleup)^{-1/2}\grad\btimes$ \cite{Moffat, Moses}.\footnote{Let 
us make explicit how this operator is defined. Suppose that $e_n(\bx)$ are the eigenfunctions of the 
minus Laplacian $-\bigtriangleup$ in the domain $\Lambda$. With appropriate boundary conditions
allowing integration by parts (e.g. Dirichlet), the corresponding eigenvalues $\lambda_n$ are 
nonnegative. Since the eigenfunctions form a complete set for any reasonable domain, we can 
expand any vector function $\bv \in L^2(\Lambda)$ as $\bv(\bx)=\sum_n {\bf c}_n e_n(\bx)$ for some 
vector coefficients ${\bf c}_n$. In that case, the operator is defined by $(\Sigma \bv)(\bx)
= \sum_n \grad e_n(\bx)\btimes{\bf c}_n/\sqrt{\lambda_n}$. This gives a constructive 
definition of the operator that could be used in practical calculations.} Acting on solenoidal 
vector fields (such as an incompressible velocity field), this operator satisfies $\Sigma^2=I$. 
Hence, its eigenvalues are $\pm 1$. We can therefore write $\bv=\bv^+ + \bv^-$, with $\Sigma\bv^\pm = 
\pm\bv^\pm.$ In fact, formally $\bv^\pm = P^\pm\bv,$ where $P^\pm={{1}\over{2}}(1\pm\Sigma)$ are 
the projectors onto $\pm$ helical subspaces. Likewise, we can decompose the vorticity field as $\bom=
\bom^+ - \bom^-$, for $\Sigma\bom^\pm = \pm\bom^\pm$. Notice our sign convention for defining $\bom^-$. 
Thus, $\bom^\pm =\pm \grad\btimes\bv^\pm$. If the fluid flow is contained in a periodic box, 
then these decompositions agree with those obtained earlier by expansion into helical waves. 

Because the $+$ and $-$ helical modes belong to distinct eigenspaces of self-adjoint operator $\Sigma$,
they are always orthogonal. It follows that the quadratic invariants of energy and helicity may be 
written as $E=E^++E^-$ and $H=H^+-H^-$, where
\be E^\pm(t)={{1}\over{2}}\int d^3\bx\,\,|\bv^\pm(\bx,t)|^2,\,\,\,\,
                     H^\pm(t)=\int d^3\bx \,\,\bom^\pm(\bx,t)\bdot\bv^\pm(\bx,t) \lb{real-EH-Hdecomp} \ee
Note in particular that 
\be H^\pm(t)=\langle\bv^\pm,(-\bigtriangleup)^{1/2}\bv^\pm\rangle\geq 0. \lb{pos-H-pm} \ee
Thus, the partial helicities $H^\pm$ are always nonnegative (with our conventions). 

\subsection{The Helical Decomposition for Filtered Quantities}

As we have just shown, the helical decomposition can be made in any domain 
intrinsically and its applicability is not restricted to flow in a periodic box. We consider 
here the dynamics in a general domain, without use of Fourier analysis. Additional 
insight can, in fact, be obtained by considering the transfer dynamics of helical modes 
in physical space. However, to discuss transfer, we must distinguish different scales of motion.
To resolve the dynamics simultaneously in space and in scale, we shall employ the same 
{\it filtering approach} that is used in large-eddy simulation \cite{MK}. In this approach
a low-pass filtered velocity $\vL(\bx,t)$ with scales $<\ell$ removed is introduced by a
convolution $\vL=G_\ell*\bv$ with a smooth filtering function $G_\ell(\bx)=\ell^{-3}G(\bx/\ell)$. 
The filtered velocity obeys the equation
\be \partial_t\vL(\bx,t)+ \grad\bdot[\vL(\bx,t)\vL(\bx,t)+\btau(\bx,t)] = 
                         \fL(\bx,t)-\grad\pL(\bx,t)+\nu\bigtriangleup\vL(\bx,t)  \lb{LES} \ee
Here $\fL,\pL$ are the filtered forcing and pressure, respectively, and $\btau=\overline{(\bv\bv)}_\ell
-\vL\vL$ is the {\it turbulent stress}, or spatial momentum transport induced by the 
eliminated small-scale turbulence. We may similarly consider a filtered vorticity
$\overline{\bom}_\ell(\bx,t)$ and its corresponding equation (which is most simply obtained by 
taking the curl of (\ref{LES}) above). The total energy in the large-scales $>\ell$ is represented 
in terms of filtered quantities by 
\be E(\ell,t)= {{1}\over{2}}\int d\bx \,\,|\overline{\bv}_\ell(\bx,t)|^2. \lb{EL} \ee
and the total helicity in the large-scales $>\ell$ by
\be H(\ell,t)= \int d\bx \,\,\overline{\bv}_\ell(\bx,t)\cdot\overline{\bom}_\ell(\bx,t). \lb{HL} \ee
Note that these quantities represent cumulative values in the large-scales, {\it not} the analogues 
of Fourier spectra but instead spectra integrated in wavenumber from zero to $\sim 1/\ell$. 

We may assume the filtering operation to commute with $\Sigma$ and $P^\pm$ (e.g. choose $G_\ell=
e^{\ell^2\bigtriangleup}$). It follows that $\overline{(\bv^\pm)}_\ell=\vLpm.$ Hence, we may introduce 
without any ambiguity the helical decompositions of large-scale velocity, $\vL=\overline{\bv}_\ell^+ 
+\overline{\bv}_\ell^-$, and of large-scale vorticity, $\oL=\overline{\bom}_\ell^+ -\overline{\bom}_\ell^-.$
The energy contained in $\pm$ modes at length-scales $>\ell$ is 
\be E^\pm (\ell,t)= {{1}\over{2}}\int d\bx \,\, |\overline{\bv}_\ell^\pm(\bx,t)|^2 \lb{pm-EL} \ee
so that $E(\ell,t) = E^+(\ell,t)+E^-(\ell,t)$. Likewise, the helicity contained in $\pm$ modes 
at length-scales $>\ell$ is
\be H^\pm (\ell,t)= \int d\bx \,\, \overline{\bv}_\ell^\pm(\bx,t)
                                \cdot\overline{\bom}_\ell^\pm(\bx,t). \lb{pm-HL} \ee
so that $H(\ell,t) = H^+(\ell,t)-H^-(\ell,t)$. As we have defined these quantities, $E^\pm (\ell,t)\geq 0$
and $H^\pm (\ell,t)\geq 0$ for both $\pm$ helical modes. 

It is not difficult using the Navier-Stokes dynamics to obtain the evolution equations of large-scale
energy and helicity. For example, for energy one easily calculates that 
\be {{dE^\pm}\over{dt}}(\ell,t) = F^\pm_E(\ell,t)-\Pi^\pm_E(\ell,t)\pm R_E(\ell,t)
                                                -D^\pm_E(\ell,t) \lb{pm-LE-eq} \ee
where 
\be F^\pm_E(\ell,t)= \int d\bx \,\, \overline{\bF}_\ell^\pm(\bx,t)\bdot \overline{\bv}_\ell^\pm(\bx,t) 
                                                                  \lb{pm-LEF} \ee
is the input of energy in the $\pm$ modes at large length-scales $>\ell$, 
\be D^\pm_E(\ell,t) = \nu \int d\bx \,\, |\grad\overline{\bv}_\ell^\pm(\bx,t)|^2  \lb{pm-LED} \ee
is the dissipation of energy in the $\pm$ modes at length-scales $>\ell$,   
\be \Pi^\pm_E(\ell,t) = -\int d\bx \,\, \nabla\overline{\bv}_\ell^\pm(\bx,t)\bdots
                                            \btau(\bx,t)  \lb{pm-Eflux} \ee
is the net flux of energy from the $\pm$ modes at length-scales $>\ell$ into the 
subgrid length-scales $<\ell$, and, finally, 
\be R_E(\ell,t) = \pm \int d\bx \,\, \vLmp\bdot (\vL\bdot\grad)\vLpm \lb{pm-LER} \ee
is the transfer of energy entirely in the large length-scales between $+$ and $-$ components. 
We have already studied in \cite{CCEH} the statistics of the flux quantities $\Pi^\pm_E(\ell,t).$
The formula (\ref{pm-LER}) gives some further insight into the mechanism of transfer of energy 
between $+$ and $-$ modes. It is seen that a $\pm$-polarized helical wave, when advected by 
a mode of any other polarity, will not stay pure $\pm$-polarized but will develop a component 
of the opposite sign polarity. This is the essential process contributing to $R_E(\ell,t).$ 

Likewise, for helicity, it is not difficult to show that 
\be {{dH^\pm}\over{dt}}(\ell,t) = F^\pm_H(\ell,t)-\Pi^\pm_H(\ell,t)+R_H(\ell,t)
                                                -D^\pm_H(\ell,t) \lb{pm-LH-eq} \ee
where 
\be F^\pm_H(\ell,t)= 2\int d\bx \,\, \overline{\bF}_\ell^\pm(\bx,t)\bdot \overline{\bom}_\ell^\pm(\bx,t) 
                                                                      \lb{pm-LHF} \ee
is the input of helicity in the $\pm$ modes at large length-scales $>\ell$, 
\be D^\pm_H(\ell,t) = 2\nu \int d\bx \,\, \grad\overline{\bv}_\ell^\pm(\bx,t)\bdots
                                  \grad\overline{\bom}_\ell^\pm(\bx,t) \lb{pm-LHD} \ee
is the dissipation of helicity in the $\pm$ modes at length-scales $>\ell$, 
\be \Pi^\pm_H(\ell,t) = -2\int d\bx \,\, \grad\overline{\bom}_\ell^\pm(\bx,t)\bdots
                                            \btau(\bx,t)  \lb{pm-Hflux} \ee
is the net flux of helicity from the $\pm$ modes at length-scales $>\ell$ into the 
subgrid length-scales $<\ell$, and, finally, 
\be R_H(\ell,t) = \int d\bx \,\,[ 
          \vLmp\bdot (\vL\bdot\grad)\oLpm+ \vLpm\bdot(\vL\bdot\grad)\oLmp 
          \pm \vLpm\bdot (\oL\bdot\grad)\vLmp ]. \lb{pm-LHR} \ee
Integration by parts shows that the latter expression is the same for both $\pm$ signs. 
It represents a transfer of helicity entirely in the large length-scales between $+$ and
$-$ components. We see that there are two mechanisms of such transfer. First, as for energy,
advection of a $\pm$-polarized helical wave will produce a component of the opposite sign 
and this gives a transfer of helicity between $+$ and $-$ modes. In addition, a $\pm$-polarized 
helical wave which stretches a vortex of any polarity will generate vorticity of the opposite 
$\mp$ polarity. This mechanism of transfer of helicity between $+$ and $-$ modes is represented 
by the third term in (\ref{pm-LHR}). The vortex-stretching mechanism for transfer between 
$+$ and $-$ modes was already considered by Waleffe \cite{Waleffe}, who showed that it 
is the only mechanism that survives in the limit of large scale separation. 

\subsection{Relation Between Filtered and Spectral Quantities}

The quantities defined in the filtering approach above, when ensemble-averaged and with $\ell$ set $=0$, 
coincide with the corresponding total integrated spectral quantities defined in Section 2.3. Thus, 
from (\ref{pm-LE-eq}),(\ref{pm-LH-eq}) by taking the limit $\ell\rightarrow 0$ one rederives 
the equations (\ref{tot-pm-E-balance}),(\ref{tot-pm-H-balance}). Of course, in that limit 
the fluxes $\Pi^\pm_E(\ell,t),\Pi^\pm_H(\ell,t)$ must vanish, for $\nu>0$. From this 
rederivation of (\ref{tot-pm-E-balance}),(\ref{tot-pm-H-balance}), we obtain interesting 
identities for integrated transfers:
\be \int_0^\infty dk\,\, {{T_H(k,t)}\over{4k}} = 
    \pm \langle\bv^\mp\bdot (\bv\bdot\grad)\bv^\pm\rangle, \lb{RE-identity} \ee
with both sides equal to $R_E$, and 
\be \int_0^\infty dk\,\, kT_E(k,t) = 
    \langle \bv^\mp\bdot (\bv\bdot\grad)\bom^\pm+ \bv^\pm\bdot(\bv\bdot\grad)\bom^\mp 
            \pm \bv^\pm\bdot (\bom\bdot\grad)\bv^\mp \rangle,           \lb{RH-identity} \ee
with both sides equal to $R_H$. 

For $\ell>0$, the equations (\ref{pm-LE-eq}),(\ref{pm-LH-eq}) are essentially the same as 
the spectral balance equations (\ref{pm-E-balance}),(\ref{pm-H-balance}) integrated in wavenumber 
from zero to $\sim 1/\ell$. Clearly, it is the combined expression 
\be \Pi^{\pm,<}_E(\ell,t) := \Pi^\pm_E(\ell,t)\pm R_E(\ell,t)  \lb{pm-lt-filt-Eflux} \ee
which corresponds to the flux quantity $\Pi_E^{\pm,<}(k,t)$ that we defined spectrally. 
Likewise it is
\be \Pi^{\pm,<}_H(\ell,t) := \Pi^\pm_H(\ell,t)-R_H(\ell,t).  \lb{pm-lt-filt-Hflux} \ee
which corresponds to $\Pi_H^{\pm,<}(k,t).$ Just as for the spectrally defined quantities,
$\Pi^{\pm,<}_X(\ell,t)$ represent the flow of $X$ out of $\pm$ modes for the length-scales 
$>\ell,$ with $X=E,H$. There are two mechanisms that contribute to this flux. In the first,
represented by $\Pi^\pm_X(\ell,t),\,\,X=E,H$, the conserved quantity $X$ leaves the $\pm$ 
modes at length-scales $>\ell$ and enters the subgrid modes with length-scales $<\ell$. 
In the second, represented by $R_X(\ell,t),\,\,X=E,H$, the conserved quantity $X$ leaves the $\pm$ 
modes at length-scales $>\ell$ by being transferred to opposite polarity $\mp$ modes still
at length-scales $>\ell$. The fluxes $\Pi_E^{\pm,<}(\ell,t),\Pi_H^{\pm,<}(\ell,t)$ and the similar 
spectral quantities represent the net effect of both mechanisms. 

\subsection{Kolmogorov Scaling Theory}

The same Kolmogorov scaling arguments made in Section 2.4 in wavenumber space apply also
to the filtered quantities. Hence, in the driven steady-state, the fluxes $\Pi^{\pm,<}_E(\ell,t),
\Pi^{\pm,<}_H(\ell,t)$ must have averages which are constant in the inertial-range:
\be \langle  \Pi^{\pm,<}_E(\ell) \rangle \approx F^\pm_E, \lb{pm-lt-filt-Eflux-avrg} \ee
and 
\be \langle  \Pi^{\pm,<}_H(\ell) \rangle \approx F^\pm_H, \lb{pm-lt-filt-Hflux-avrg} \ee
as long as the mean dissipations $\langle D^\pm_E(\ell)\rangle,\langle D^\pm_H(\ell)\rangle$ 
can be neglected. This will hold for all $L\gg\ell\gg \eta=1/k_E$ in the case of $\Pi^{\pm,<}_E(\ell)$
and for all $L\gg\ell\gg \xi=1/k_H$ in the case of $\Pi^{\pm,<}_H(\ell).$

However, the separate terms $\Pi^\pm_E(\ell,t)$ and $R_E(\ell,t)$ should {\it not} be expected 
to have constant means. The average $\langle R_E(\ell,t)\rangle$ will grow in magnitude through the 
inertial range, from near zero at $\ell\approx L$ to the value $R_E$ at very small $\ell$. For a very 
long inertial range, this should have value $R_E\approx {{1}\over{2}}(F_E^--F_E^+)$, as required to bring 
the energy in $+$ and $-$ modes into balance. The flux averages $\langle\Pi_E^\pm(\ell)\rangle$ will change 
from $\approx F^\pm_E$ for $\ell\approx L$ near the forcing scale to the value $F^\pm_E \pm R_E =D^\pm_E$ 
for $\ell$ approaching the dissipation range. At high enough Reynolds number, the
latter equals ${{1}\over{2}}(F_E^++F_E^-)={{\varepsilon}\over{2}}$ and, deep in the inertial range,
the $+$ and $-$ modes should each carry about half of the total energy flux. Although 
$\langle\Pi_E^\pm(\ell)\rangle$ are not constant for $\ell$ moving through the inertial range, 
the net energy flux $\Pi_E=\Pi_E^++\Pi_E^-$ satisfies 
\be     \langle\Pi_E(\ell)\rangle \approx \varepsilon  \lb{filt-Eflux-const} \ee
for {\it all} $\ell$ in the inertial range, where $\varepsilon= F^+_E+F^-_E.$
Only the fractions carried by $\pm$ modes changes. 

Just as for energy, the separate terms $\Pi^\pm_H(\ell,t)$ and $R_H(\ell,t)$ 
are not expected to have constant means. In fact, for $\ell\ll L$, one expects  
\be \langle\Pi_H^\pm(\ell)\rangle \sim {{2\langle\Pi_E^\pm(\ell)\rangle}\over{\ell}}. 
                                                        \lb{pm-filt-Hflux-avrg} \ee
Such a result follows by the same reasoning that led Kraichnan to conclude for 2D that an 
energy flux at length-scale $\ell$ would imply an enstrophy flux bigger by a factor $\ell^{-2}$
at small-scales. This helicity flux continuously grows as $\ell$ decreases through the inertial range, 
from a value $\approx F_H^\pm$ for $\ell\approx L$ and eventually matches onto the dissipation $D_H^\pm 
\sim 2D_E^\pm/\eta\propto \varepsilon/\eta$ at $\ell\approx \eta$. See (\ref{K41-pm-H-dissipation}). 
Of course, the growing term must cancel in the total flux---given by the difference $\Pi_H^{\pm,<}
=\Pi_H^\pm-R_H$---which has constant mean $F^\pm_H$, so that we have 
\be \langle R_H(\ell)\rangle \sim {{2\langle\Pi_E^\pm(\ell)\rangle}\over{\ell}}-F^\pm_H
                                                               \lb{pm-filt-RH-avrg} \ee
for $\ell\ll L$ in the inertial range. This is nearly zero for $\ell\approx L$, while taking 
$\ell\approx\eta$ implies that $R_H\sim \varepsilon/\eta$. This latter relation can be verified directly 
from the expression (\ref{tot-H-transfer}). Using $T_E(k)= -{{\partial}\over{\partial k}}\Pi_E(k)$ and 
integrating by parts gives
\be R_H= \int_0^\infty dk\,\, \Pi_E(k) \lb{tot-H-transfer-Pi} \ee
By Kolmogorov scaling analysis we expect that $\Pi_E(k)\approx\varepsilon$ over a range of wavenumbers
of length $\sim k_E$. Thus, $R_H\sim \varepsilon\cdot k_E$. 

The divergence in the separate terms $\Pi_H^+,\Pi_H^-$ cancels as well in the 
difference $\Pi_H=\Pi_H^+-\Pi_H^-$, giving a constant net flux of helicity
\be      \langle\Pi_H(\ell)\rangle \approx \delta  \lb{filt-Hflux-const} \ee
for all $\ell\ll L$ in the inertial range, where $\delta=F_H^+ -F_H^-.$ This means that the 
reflection symmetry-breaking at the large scales continues to be expressed in the statistics 
of the helicity flux throughout the inertial range. However, at sufficiently high Reynolds 
number and sufficiently deep in the inertial range,  
\be \langle\Pi_H^\pm(\ell)\rangle \sim {{\varepsilon}\over{\ell}} 
                                                        + F^\pm_H.  \lb{deep-pm-filt-Hflux-avrg} \ee
Therefore, parity-symmetry is restored to leading order.

\subsection{Simulation Results}

We have checked the predictions in the preceding sections in our numerical simulation, as previously 
described. We employed a Gaussian filter, given in Fourier space as $\widehat{G}_{k_c}(\mathbf{k})=
e ^{-(\pi^2 \bk\bdot\bk/24 k^2_c)}$ with $k_c=\pi/\ell$. Here, $k_c$ is the cutoff 
wavenumber and $\ell$ is the filter width. 

Our first result is the averages of the total filtered energy and helicity fluxes defined by 
$\Pi_E(\ell,t)=\Pi_E^+(\ell,t)+\Pi_E^-(\ell,t)$ and $\Pi_H(\ell,t)=\Pi_H^+(\ell,t)-\Pi_H^-(\ell,t)$.
These are plotted in Fig.~7, normalized by the total inputs $\epsilon$ and $\delta$ respectively. 
A short plateau between $30 \leq \ell/\eta \leq 80$ indicates these filtered fluxes have a small inertial range. 

Let us now consider the fluxes in the $\pm$ helical channels, which we have argued not to be constant in the same range of scales. 
We begin with the energy. From the values reported earlier in Table 1, we expect that $\langle\Pi^+_E(\ell)\rangle/\varepsilon$ 
should decrease going to small scales through the 
inertial range, from $F_E^+/\varepsilon\approx 0.90$ at large scales to $D_E^+/\varepsilon\approx 0.53$
at the beginning of the dissipation range. On the other hand, $\langle\Pi^-_E(\ell)\rangle/\varepsilon$ 
should increase through the inertial range, from $F_E^-/\varepsilon\approx 0.10$ at large scales 
to $D_E^-/\varepsilon\approx 0.47$ at the start of the dissipation range. In the far dissipation range, 
both of the fluxes should decay to zero. The results in Fig.~8 agree well with these expectations. 

\noindent No theoretical prediction has been made for the {\it rate} of change of $\langle\Pi^\pm_E(\ell)
\rangle/\varepsilon$ through the inertial range, but they appear to be reasonably well fit by power laws shown in Fig.~8:
\be \langle\Pi^+_E(\ell)\rangle/\varepsilon \sim \ell^{0.19},\,\,\,\, 
    \langle\Pi^+_E(\ell)\rangle/\varepsilon \sim \ell^{-0.50}. \lb{emp-pow} \ee
Of course, this apparent power-law behavior can hold over a limited range only, else the limits $D_E^\pm/\varepsilon$ 
at the beginning of the dissipation range could not be achieved. 

The increasing equalization of $\langle\Pi^+_E(\ell)\rangle$ and $\langle\Pi^-_E(\ell)\rangle$ 
at small length scales observed in Fig.~8 is due to the transfer from $-$ modes to $+$ modes.
It is expected that $\langle R_E(\ell)\rangle/\varepsilon$ will decrease monotonically from near zero 
at the beginning of the inertial range to a final value $\langle R_E \rangle/\varepsilon \approx -0.37$ in the far 
dissipation range (using the values in Table 1). This quantity is plotted in Fig.~9.

\noindent The expected behavior is well-confirmed, including the asymptotic numerical
value at very small scales. 

Now let us consider the helicity. It is predicted from (\ref{pm-LH-eq}) and (\ref{emp-pow}) that $\langle\Pi^+_H(\ell)\rangle/\delta$ should 
increase going to small scales through the inertial range, from $F_H^+/\delta \approx 1.43$ 
at large scales to $D_H^+/\delta\approx 16.5$ at the beginning of the dissipation range. 
In addition, $\langle\Pi^-_H(\ell)\rangle/\delta$ should also increase through the inertial range, 
from $F_H^-/\delta\approx 0.43$ at large scales to $D_H^-/\delta\approx 15.5$. Proceeding into the 
far dissipation range, both of the fluxes should decay to zero. In Fig.~10, we 
show $\langle\Pi^+_H(\ell)\rangle/\delta$ and $\langle\Pi^-_H(\ell)\rangle/\delta$ as functions of $\ell/\eta$.

\noindent Given the approximate power-law behavior observed for 
$\langle\Pi^\pm_E(\ell)\rangle/\varepsilon$, we expect also approximate power-laws for 
$\langle\Pi^\pm_H(\ell)\rangle/\delta$, because of the relation (\ref{pm-filt-Hflux-avrg}).
More specifically, using (\ref{emp-pow}), that relation predicts that 
\be \langle\Pi^+_H(\ell)\rangle/\delta \sim \ell^{-0.81},\,\,\,\,
    \langle\Pi^+_H(\ell)\rangle/\delta \sim \ell^{-1.5}. \lb{new-pow} \ee
From the plots shown in Fig.~10, we see the numerical data are consistent with the above
power law relations.
In a very long inertial range it is expected (\ref{deep-pm-filt-Hflux-avrg}) is valid for both of these 
quantities, which should scale as $\sim \ell^{-1}$. The data obtained indicates that our simulation does not reach such an asymptotic limit. 

The average helicity transfer between $\pm$ modes, $\langle R_H(\ell)\rangle/\delta$, is expected 
to increase from near zero to $\langle R_H \rangle/\delta\approx 
15.1$ from large scales to small scales based on Table 1. The numerical results presented in Fig.~11:

\noindent As shown in Fig.~11, a power law $\langle R_H(\ell)\rangle/\delta
\sim \ell^{-1.5}$ can be observed. Again, 
if a very long inertial range exists the exponent should be -1 not -1.5. From Fig.~11, we obtain $\langle R_H\rangle/\delta \approx 11.22 $ in the far dissipation range which is somewhat smaller than expected. This discrepancy is possibly attributed to the ``smearing'' effect of the filter.

Unlike the above quantities with increasing and decreasing behaviors, the fluxes $\langle\Pi_E^{\pm,<}(\ell)\rangle$ and $\langle\Pi_H^{\pm,<}(\ell)\rangle$ are expected to have averages that are constant over the inertial range. 
Precisely, based on (\ref{pm-lt-filt-Eflux-avrg}), (\ref{pm-lt-filt-Hflux-avrg}) and Table 1, $\langle\Pi_E^{+,<}\rangle/\varepsilon$ should be 0.90 in the inertial range and 0.37 in the 
far dissipation range, while $\langle\Pi_E^{-,<}\rangle/\varepsilon$ should be 0.10 in the inertial range 
and -0.37 in the far dissipation range; similarly, $\langle\Pi_H^{+,<}\rangle/\delta$ should be 1.43
in the inertial range and $\langle\Pi_H^{-,<}\rangle/\delta$ should be 0.43, while both go to $-15.1$ 
in the far dissipation range. As seen in Fig.~12, these expectations are well confirmed.

\noindent Since the plateaux are short, the obtained values are somewhat smaller than expected. On the other hand, the values in the far dissipation range are quite close to those predicted. 

\section{Conclusions}

A major theoretical goal of this work was to explain the mechanism which permits a joint forward 
cascade of energy and helicity in 3D turbulence. By contrast, a dual cascade of energy and enstrophy occurs 
in 2D turbulence, with a forward cascade of enstrophy and an inverse cascade of energy. It has long been 
understood that this difference arises from the sign-indefiniteness of helicity \cite{BFLLM,Kr73}, whereas
enstrophy is nonnegative. We have argued that the contrast is seen most clearly in the helical decomposition
of the 3D flow field, used previously in \cite{DG1}. Just as has been argued in 2D for energy and enstrophy 
fluxes, in 3D a flux of energy to high wavenumber carries with it a growing flux of helicity in both $+$ and
$-$ channels. The mechanism in 3D that allows the constant fluxes of energy and helicity to coexist is the 
near cancellation of the helicity flux between the $+$ and $-$ modes. The fact that these helical-channel 
contributions almost cancel is due to the asymptotic restoration of parity symmetry at high wavenumbers. 
In fact, we have shown by very general considerations that the nonlinear transfers of energy and helicity 
between $+$ and $-$ modes will be in such a direction as to restore reflection-invariance at small 
length-scales. In addition, we have made detailed, novel predictions about the constant flux ranges in the 
$+$ and $-$ channels, both their extents in wavenumber and the plateau values achieved. We have considered 
not only spectral fluxes, appropriate to a periodic box, but also fluxes in physical space for arbitrary 
space domains using a filtering technique. The predictions have been confirmed by a numerical simulation 
of forced helical turbulence. 

The role of helicity in three-dimensional turbulence is, in our opinion, still somewhat mysterious. 
In particular, it is still unclear how energy and helicity dynamics interact in detail. The role of 
helicity in geophysical flows has been considered \cite{LET}---without being fully resolved---while its 
appearance and influence in engineering applications is still largely unexplored. We hope that this work 
will be a helpful step in the direction of better understanding the subtle manifestations of helicity 
in three-dimensional turbulence.

{\bf Acknowledgements.}
We thank Uriel Frisch, Darryl Holm, and Susan Kurien for helpful discussions. Simulations are performed 
at the Advanced Computing Laboratory at Los Alamos National Laboratory and the cluster computer supported 
by the NSF grant CTS-0079674 at the Johns Hopkins University.

\pagebreak
\vspace{0.2in}
\noindent {\bf Figure Captions}
\begin{description}
\item{
FIG.~1.  Normalized energy and helicity flux in k-space.
}                                      

\item{
FIG.~2. (a)$\Pi_E^{-,<}$, $\Pi_E^{-,>}$ and $\Pi_E$ ,(b)$\Pi_E^{+,<}$, $\Pi_E^{+,>}$ and $\Pi_E$.
}

\item{
FIG.~3. (a)$\Pi_H^{-,<}$, $\Pi_H^{-,>}$ and $\Pi_H$,(b)$\Pi_H^{+,<}$, $\Pi_H^{+,>}$ and $\Pi_H$.
}    

\item{
FIG.~4. (a)$\Pi_E^{+,>}$, $\Pi_E^{-,>}$ and (b)$\Pi_E^{+,<}$, $\Pi_E^{-,<}$. $D_E^+=0.0723$, $D_E^-=0.0685$, $R_E=-0.05152$, $F_E^+=0.1239$, $F_E^-=0.01698$.
}      

\item{
FIG.~5. (a)$\Pi_H^{+,>}$, $\Pi_H^{-,>}$ and (b) $\Pi_H^{+,<}$, $\Pi_H^{-,<}$. $D_H^+=7.30$, $D_H^-=6.86$, $R_H=6.67$, $F_H^+=0.63$, $F_H^-=0.19$. 
}  

\item{
FIG.~6. Normalized helicity flux in Helical SABRA3 model(22 shells). The inset
 shows the normalized energy($\Diamond$) and helicity fluxes($\bullet$).
}  

\item{
FIG.~7. Normalized filtered energy and helicity flux.  
} 

\item{
FIG.~8. Normalized filtered net energy fluxes in + and - modes, the solid lines are the power fitting.  
}

\item{
FIG.~9. Normalized filtered energy transfer between - and + modes.
}

\item{
FIG.~10. Normalized filtered net helicity flux in (a) + modes, (b) - modes, the solid lines are the power fitting.  
} 

\item{
FIG.~11. Normalized filtered helicity transfer between - and + modes, the solid line is the power fitting. 
} 

\item{
FIG.~12. (a) $\langle\Pi^{\pm,<}_E(\ell,t)\rangle/\epsilon$, (b)$\langle\Pi^{\pm,<}_H(\ell,t)\rangle/\delta$.
}

\end{description}  

\begin{thebibliography}{99}
\bibitem{DG1}P. D. Ditlevsen and P. Giuliani, ``Dissipation in helical turbulence,'' 
Phys. Fluids. {\bf 13}, 3508-9 (2001).
\bibitem{DG2}P. D. Ditlevsen and P. Giuliani, ``Cascades in helical turbulence,''
Phys. Rev. E {\bf 63}, 036304 (2001).
\bibitem{Mor} J. J. Moreau, ``Constantes d'un $\hat{\i}$lot tourbillonnaire en fluide parfait 
barotrope,'' C. R. Acad. Sci. Paris {\bf 252}, 2810 (1961).
\bibitem{Mof} H. K. Moffatt, ``The degree of knottedness of tangled vortex lines,''
J. Fluid Mech. {\bf 35}, 117 (1969). 
\bibitem{LET} M. Lautenschlager, D. P. Eppel, and W. C. Thacker, ``Subgrid-parameterization 
in helical flows,'' Beitr. Phys. Atmosph. {\bf 61}, 87 (1988).
\bibitem{MA} H. K. Moffat, A Tsinober,`` Helicity In Laminar And Turbulent Flow,'', Annu. Rev. Fluid Mech. {\bf 24}, 281 (1992).  
\bibitem{Kr67}R. H. Kraichnan, ``Inertial ranges in two-dimensional turbulence,'' Phys. Fluids
{\bf 10}, 1417 (1967). 
\bibitem{BFLLM} A. Brissaud, U. Frisch, J. Leorat, M. Lesieur, and A. Mazure, ``Helicity
cascades in fully developed isotropic turbulence,'' Phys. Fluids {\bf 16}, 1366 (1973).
\bibitem{Kr73}R. H. Kraichnan, ``Helical turbulence and absolute equilibrium,'' J. Fluid 
Mech. {\bf 59}, 745 (1973).
\bibitem{AL}J. C. Andr\'{e} and M. Lesieur, ``Influence of helicity on the evolution of isotropic 
turbulence at high Reynolds number,'' J. Fluid. Mech. {\bf 81}, 187 (1977).
\bibitem{BO97}V. Borue and S. A. Orszag, ``Spectra in helical three-dimensional homogeneous
isotropic turbulence,'' Phys. Rev. E {\bf 55}, 7005 (1997).  
\bibitem{CCEH}Q. Chen, S. Chen, G. L. Eyink and D. D. Holm, ``Intermittency in the joint cascade
of energy and helicity,'' submitted to Phys. Rev. Lett. (2002)
\bibitem{Waleffe}F. Waleffe, ``The nature of triad interactions in homogeneous turbulence,''
Phys. Fluids A {\bf 4}, 350 (1992).
\bibitem{Moffat}H. K. Moffat, ``Dynamo action associated with random inertial waves in a
rotating conducting fluid,'' J. Fluid Mech. {\bf 44}, 705 (1970).      
\bibitem{Moses}H. E. Moses, ``Eigenfunctions of the curl operator, rotationally
invariant
Helmholtz theorem and applications to electromagnetic theory and fluid mechanics,'' SIAM
J. Appl. {\bf 21}, 114 (1971).     
\bibitem{Lesieur}M. Lesieur, {\it Turbulence in Fluids}, 2nd edition. (Kluwer, Dordrecht, 1990). 
\bibitem{Shiyi}S. Chen and X. Shan, ``High-resolution turbulent simulations using the 
Connection Machine-2,'' Comput. Phys. {\bf 6}, 643 (1992).
\bibitem{PolShtil}W. Polifke and L. Shtilman, ``The dynamics of helical decaying turbulence,''
Phys. Fluids A {\bf 1}, 2025 (1989).
\bibitem{BifKerr}L. Biferale and R. M. Kerr, ``Role of inviscid invariants in shell models 
of turbulence,'' Phys. Rev. E {\bf 52}, 6113 (1995).
\bibitem{BBKT}R. Benzi, L. Biferale, R. M. Kerr, and E. Trovatore, ``Helical shell models 
for three-dimensional turbulence,'' Phys. Rev. E {\bf 53}, 3541 (1996). 
\bibitem{SABRA}V. S. L'vov, E. Podivilov, A. Pomyalov, I. Procaccia, and D. Vandembroucq,
``An improved shell model of turbulence,'' Phys. Rev. E {\bf 58}, 1811 (1998).
\bibitem{MK}C. Meneveau and J. Katz, ``Scale-invariance and turbulence models for large-eddy
simulation,'' Annu. Rev. Fluid Mech. {\bf 32}, 1 (2000).        

\end{thebibliography}
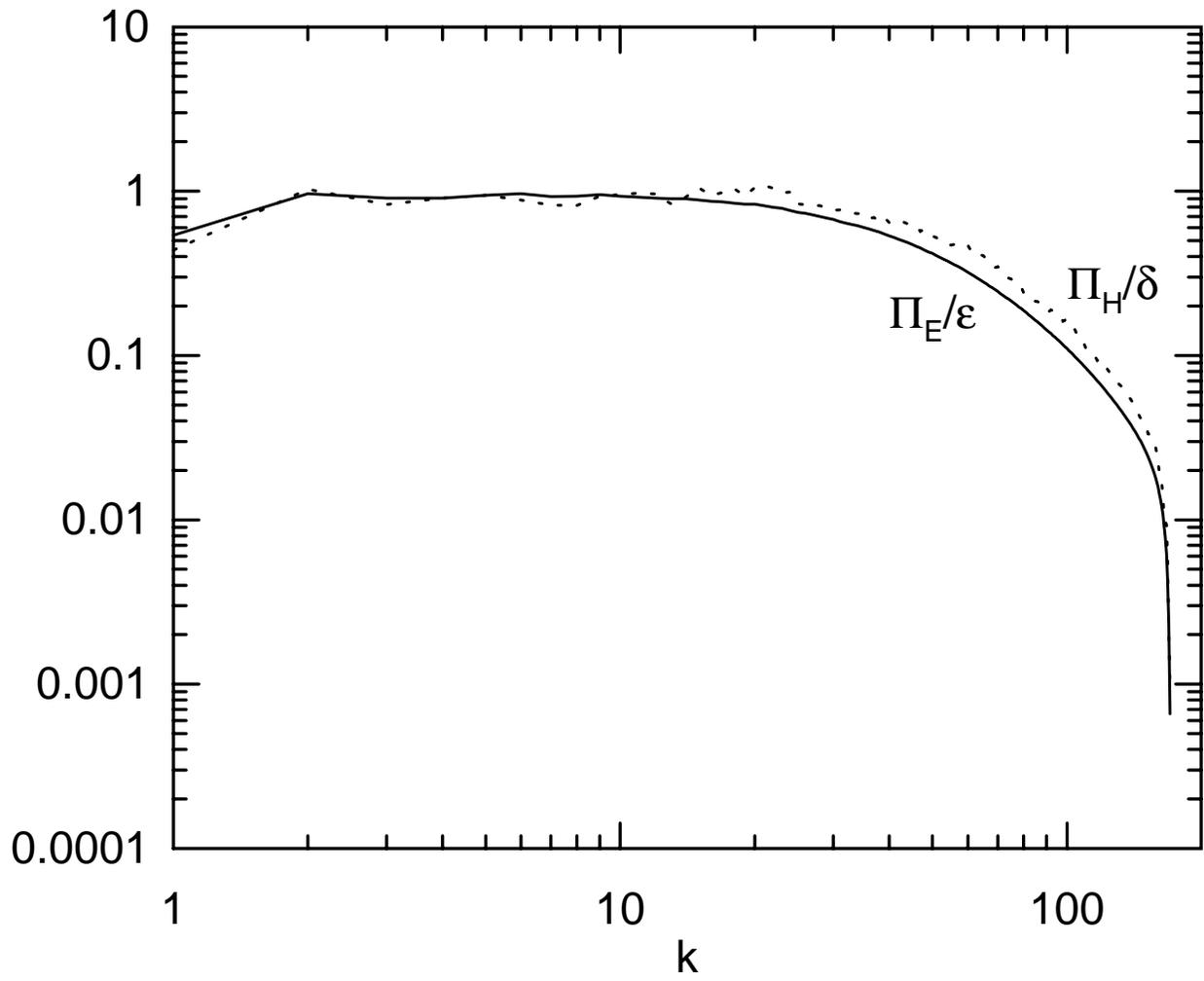
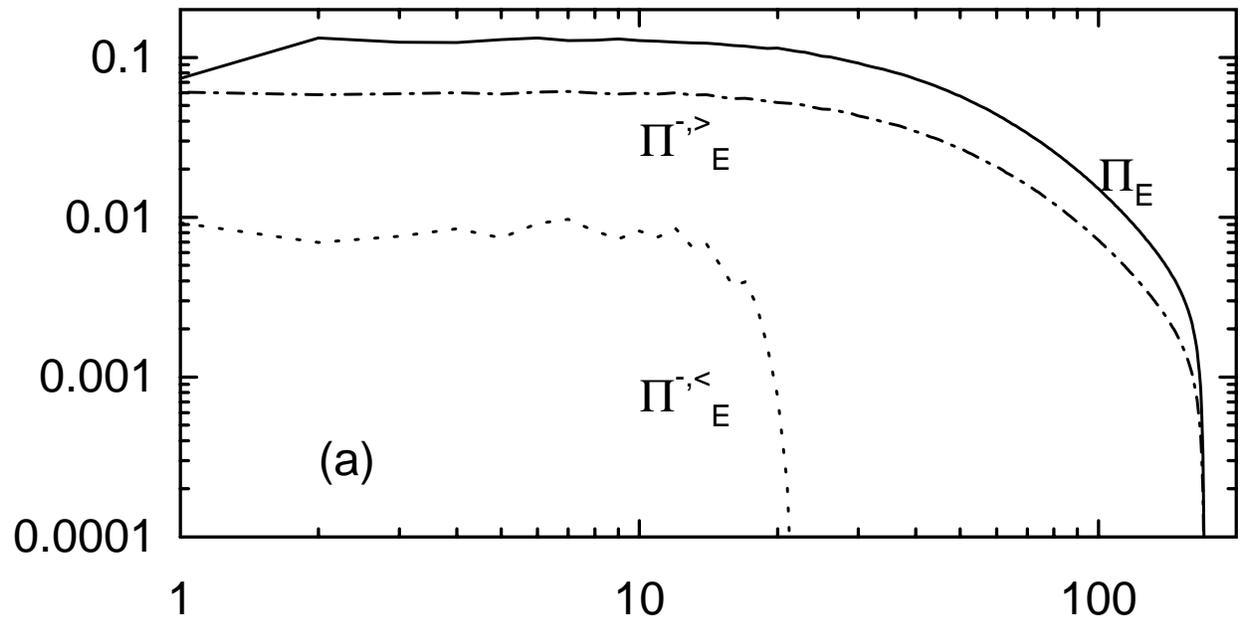
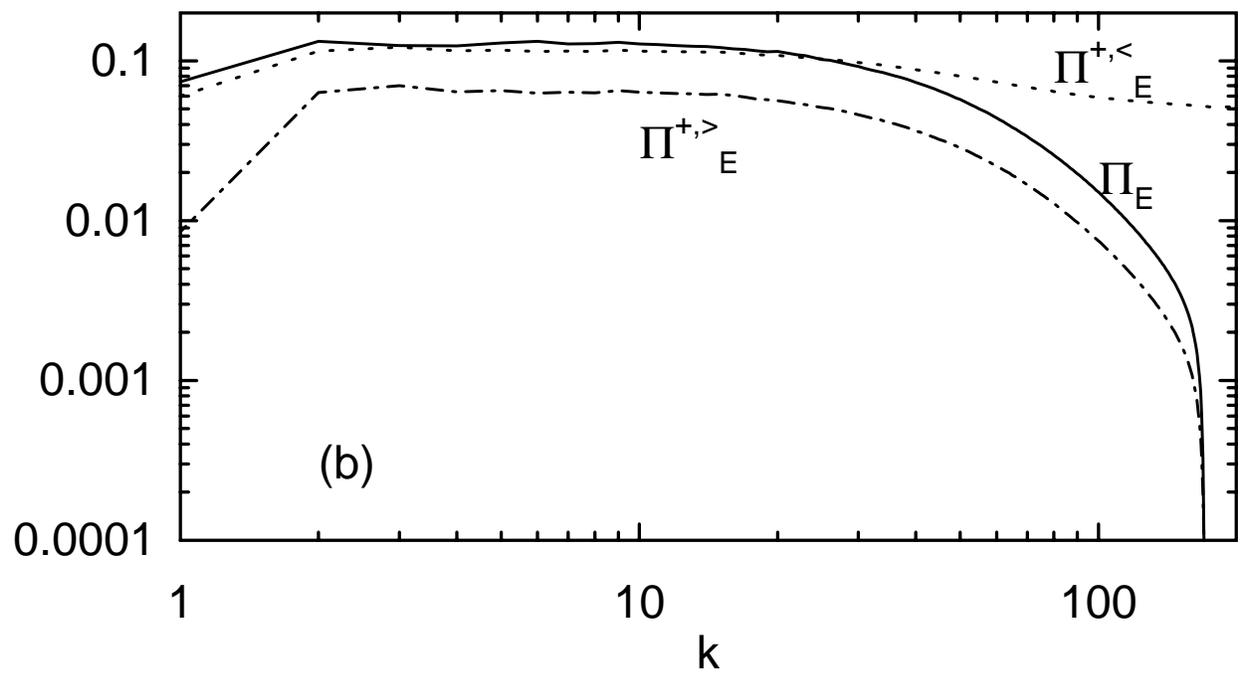
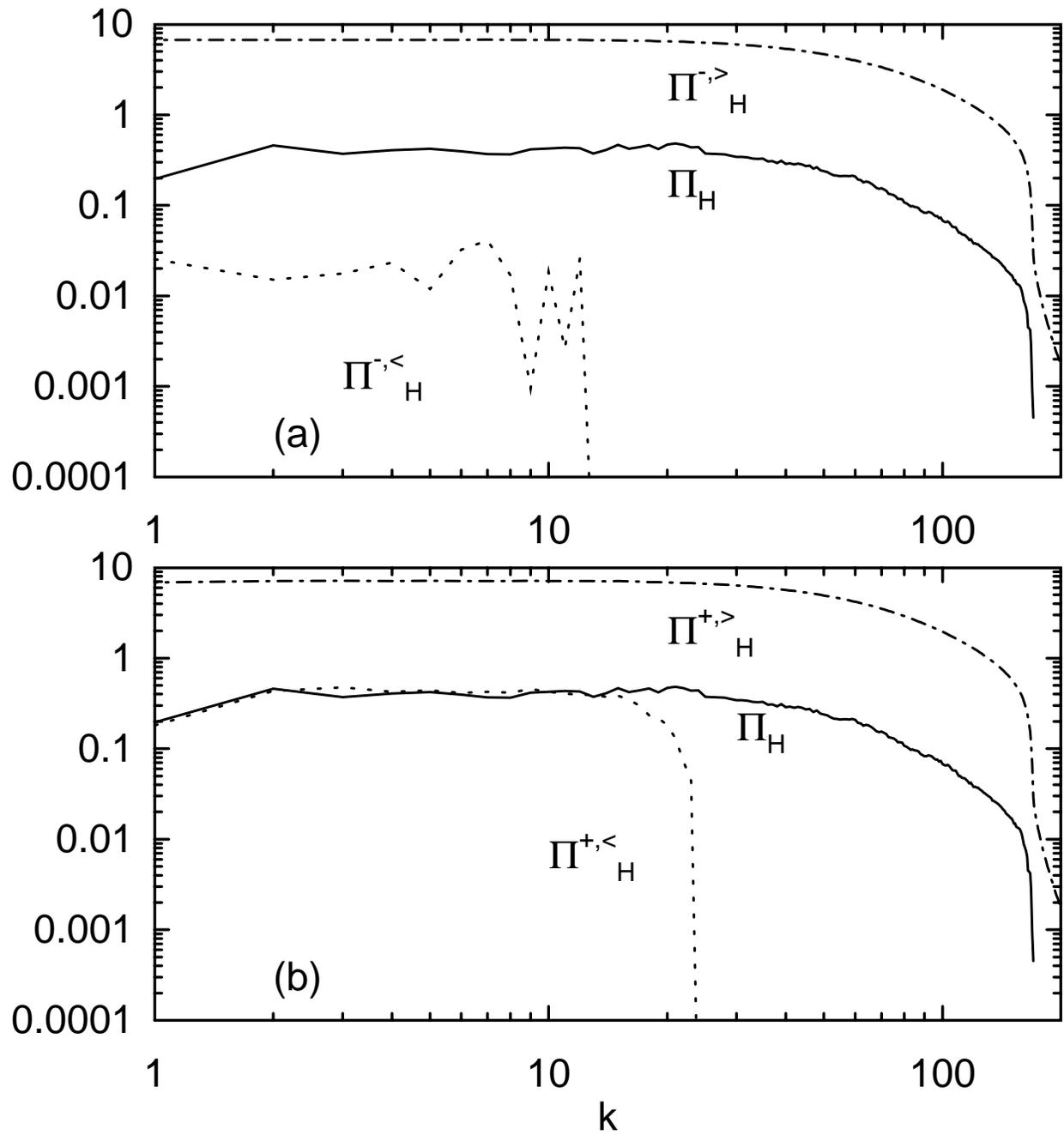
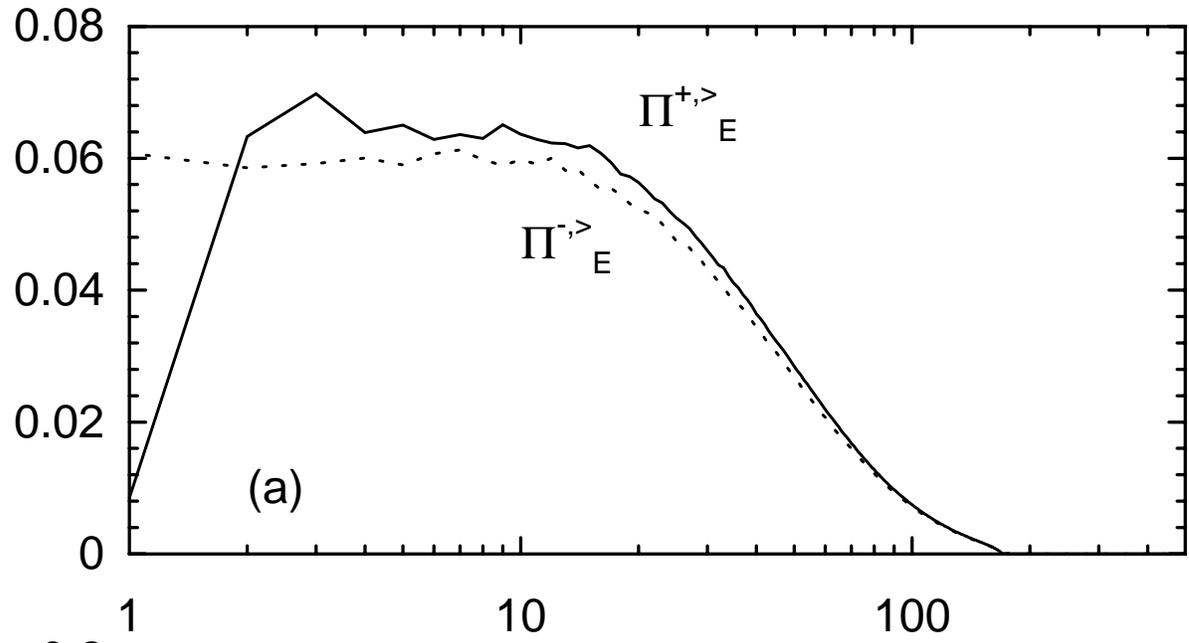
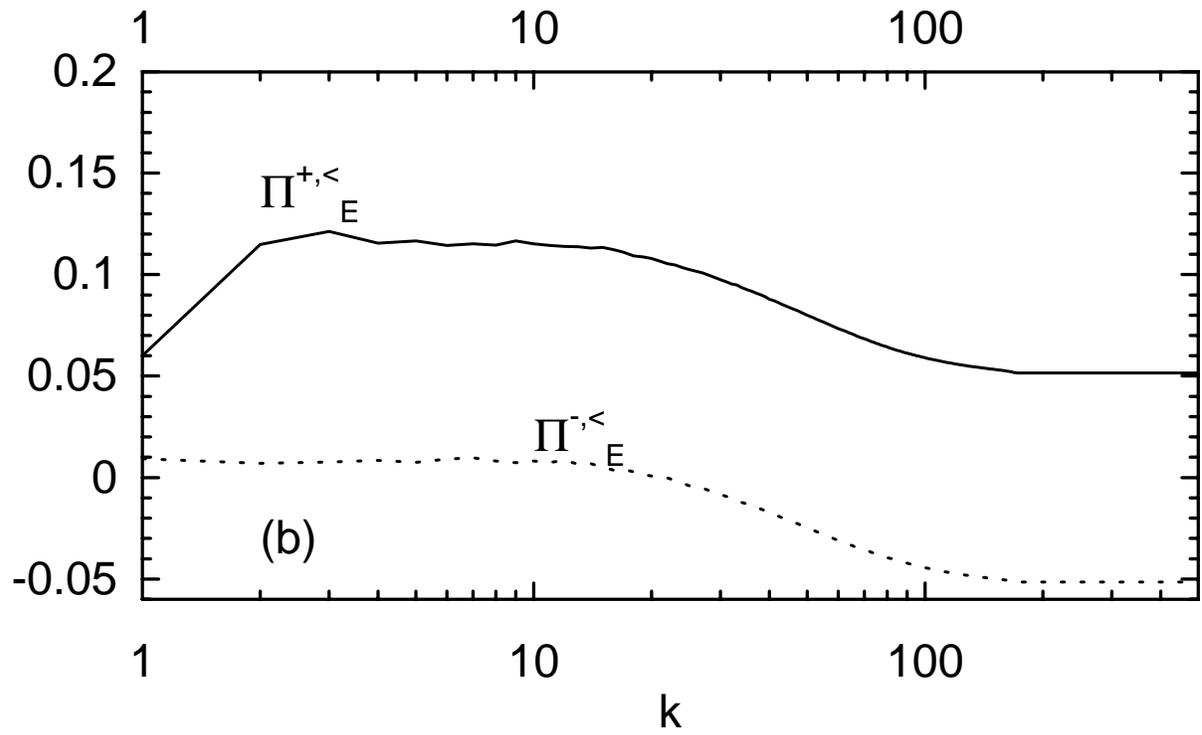
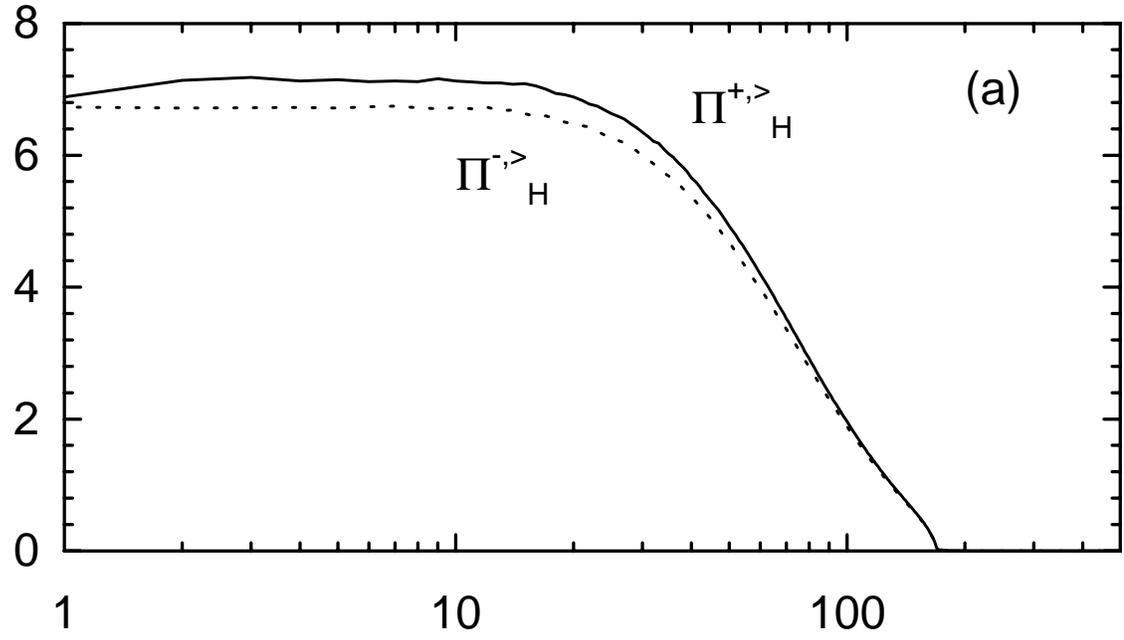
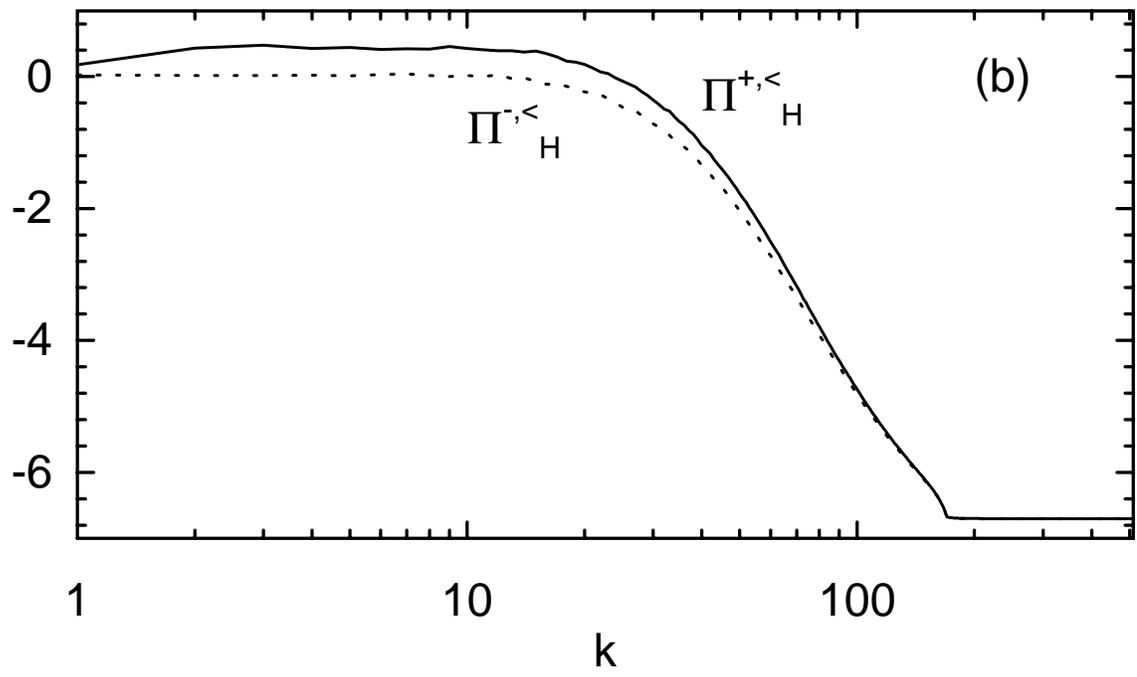
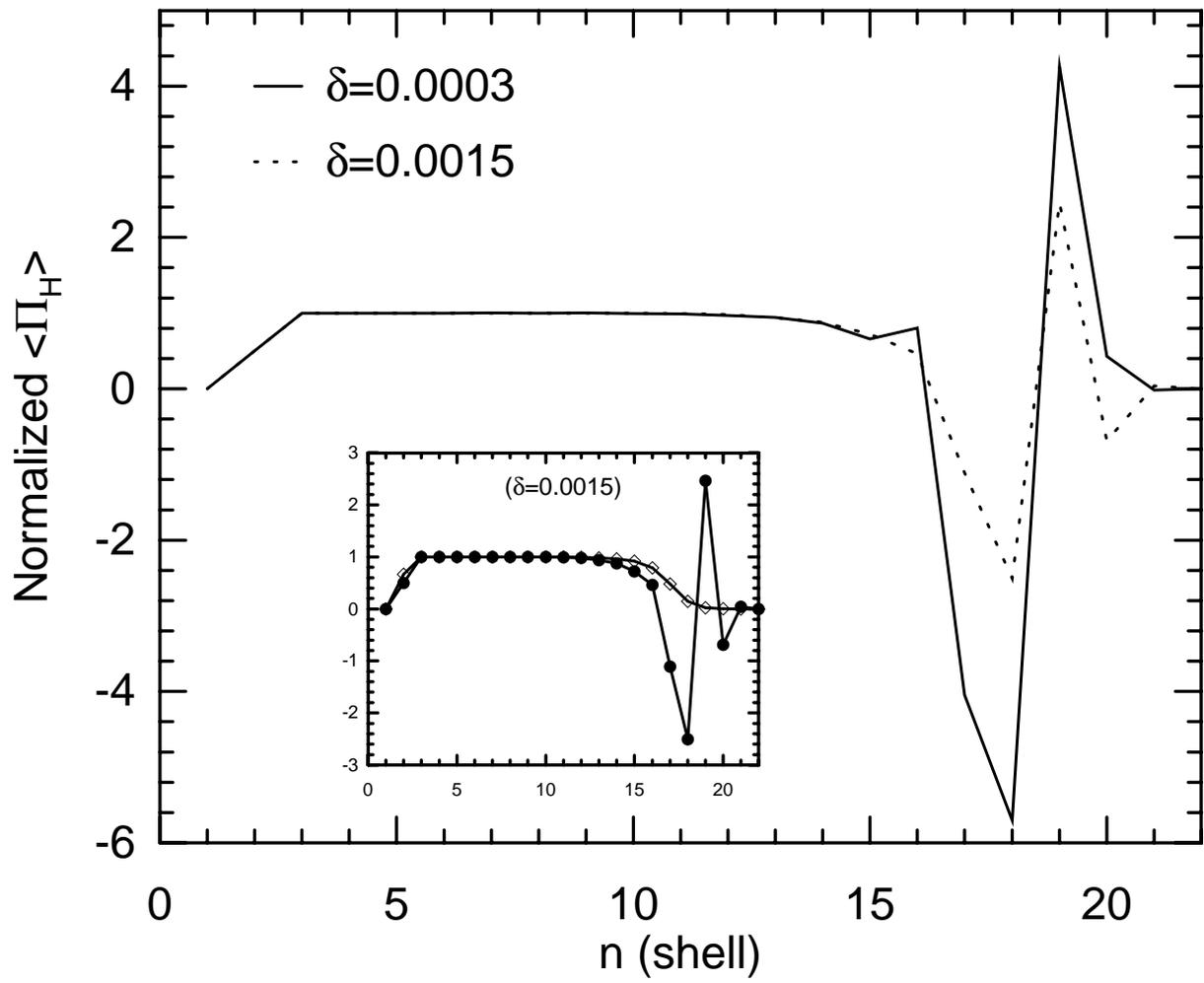
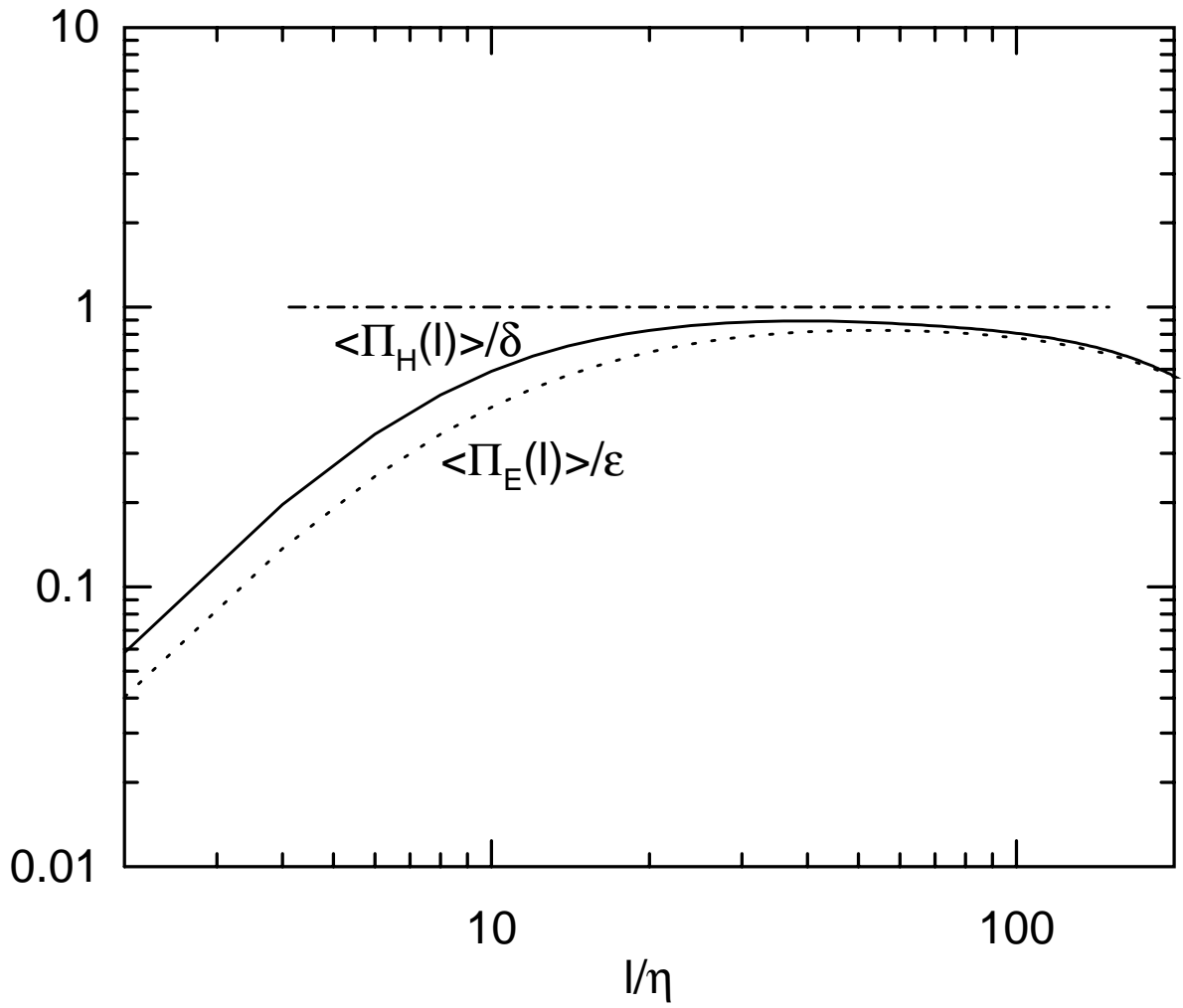
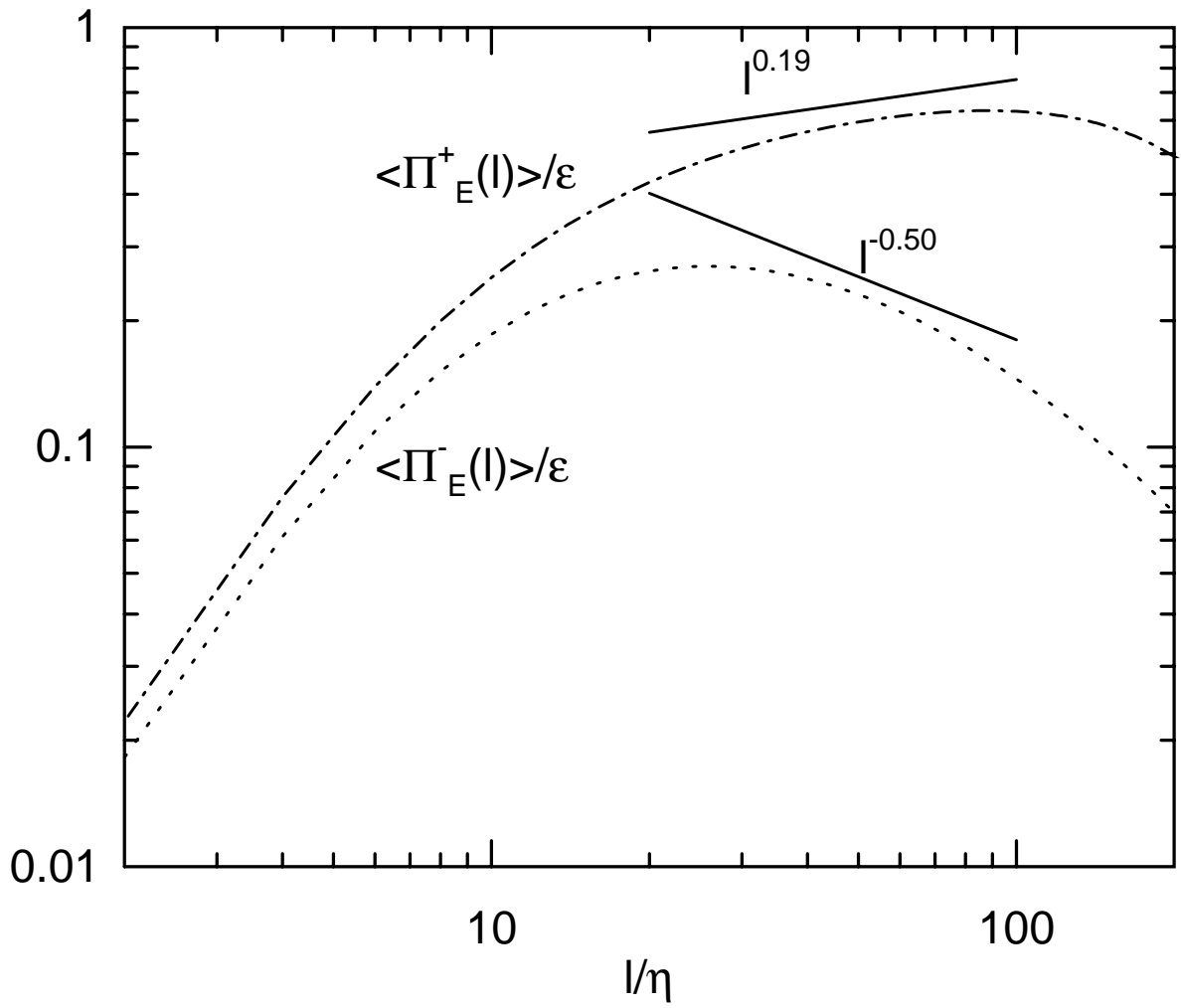
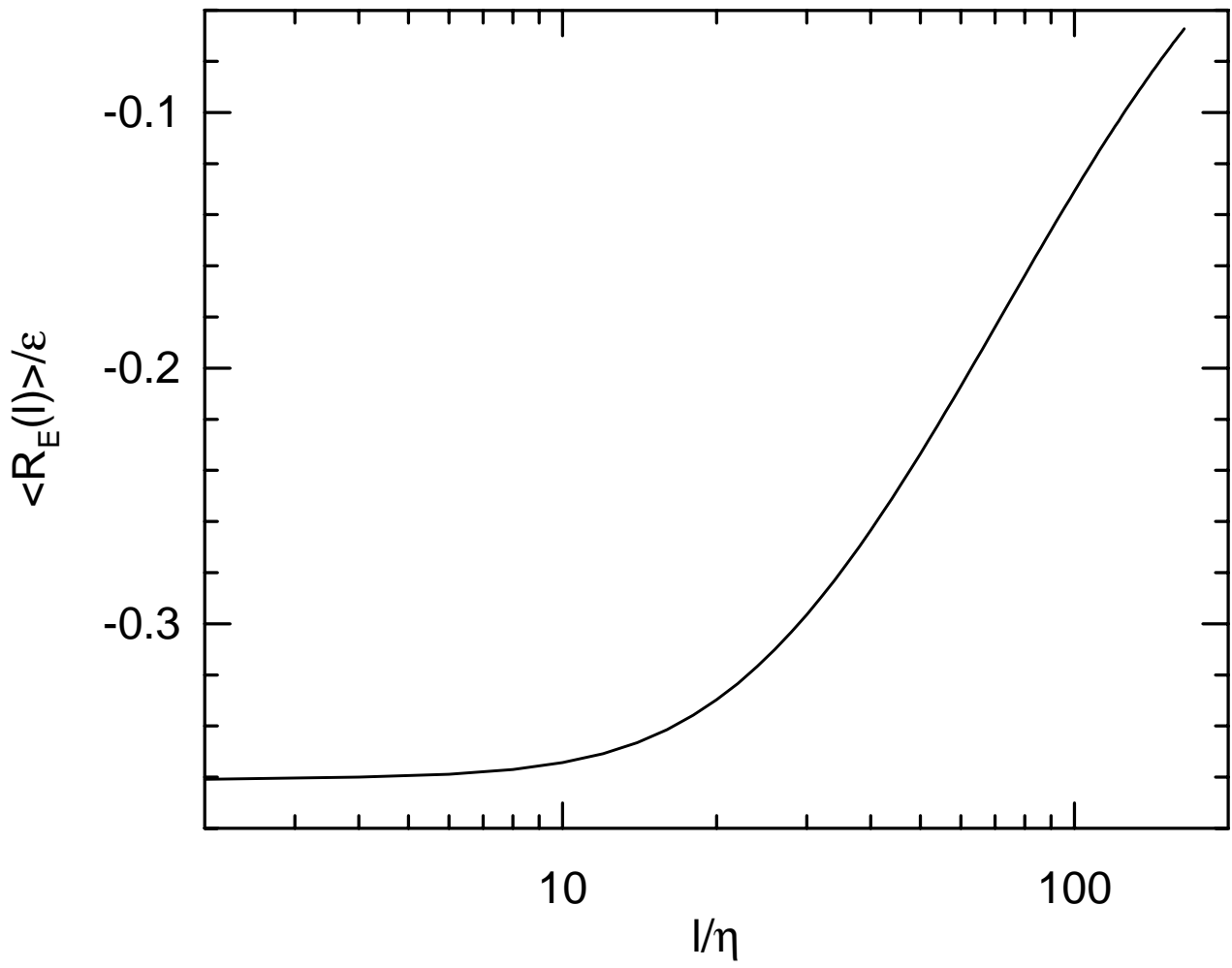
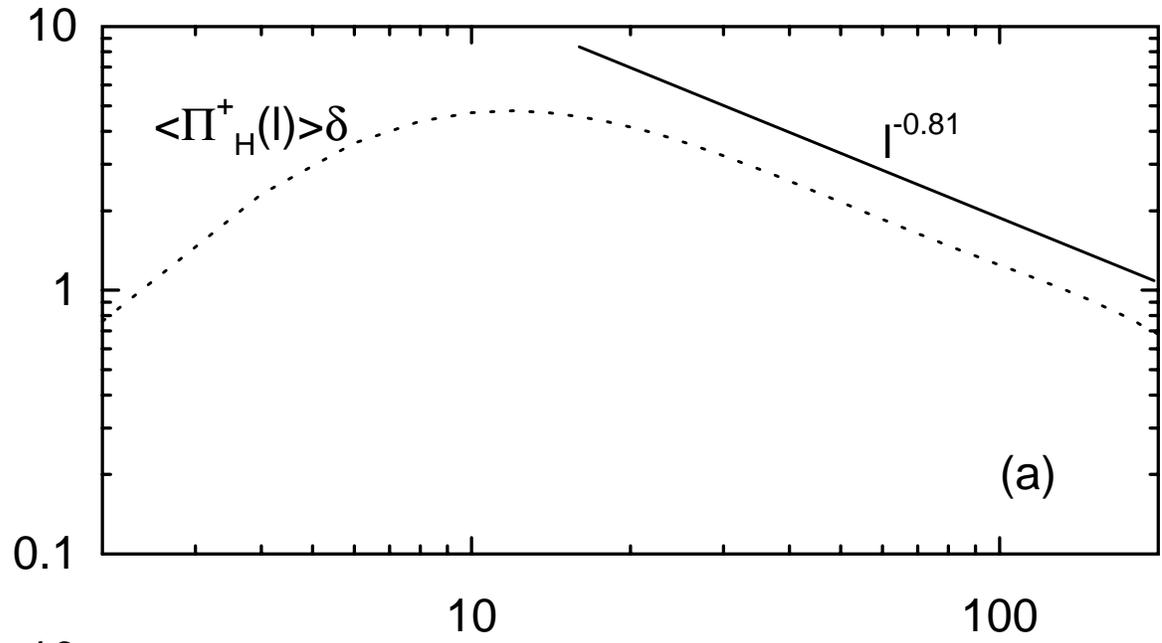
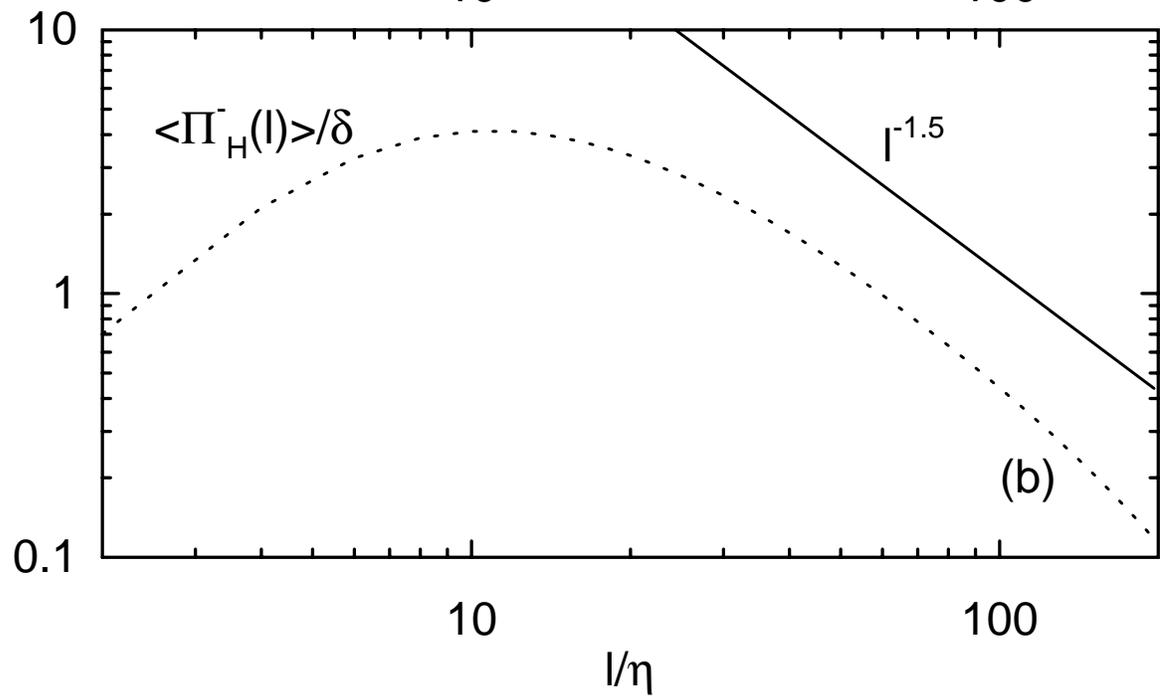
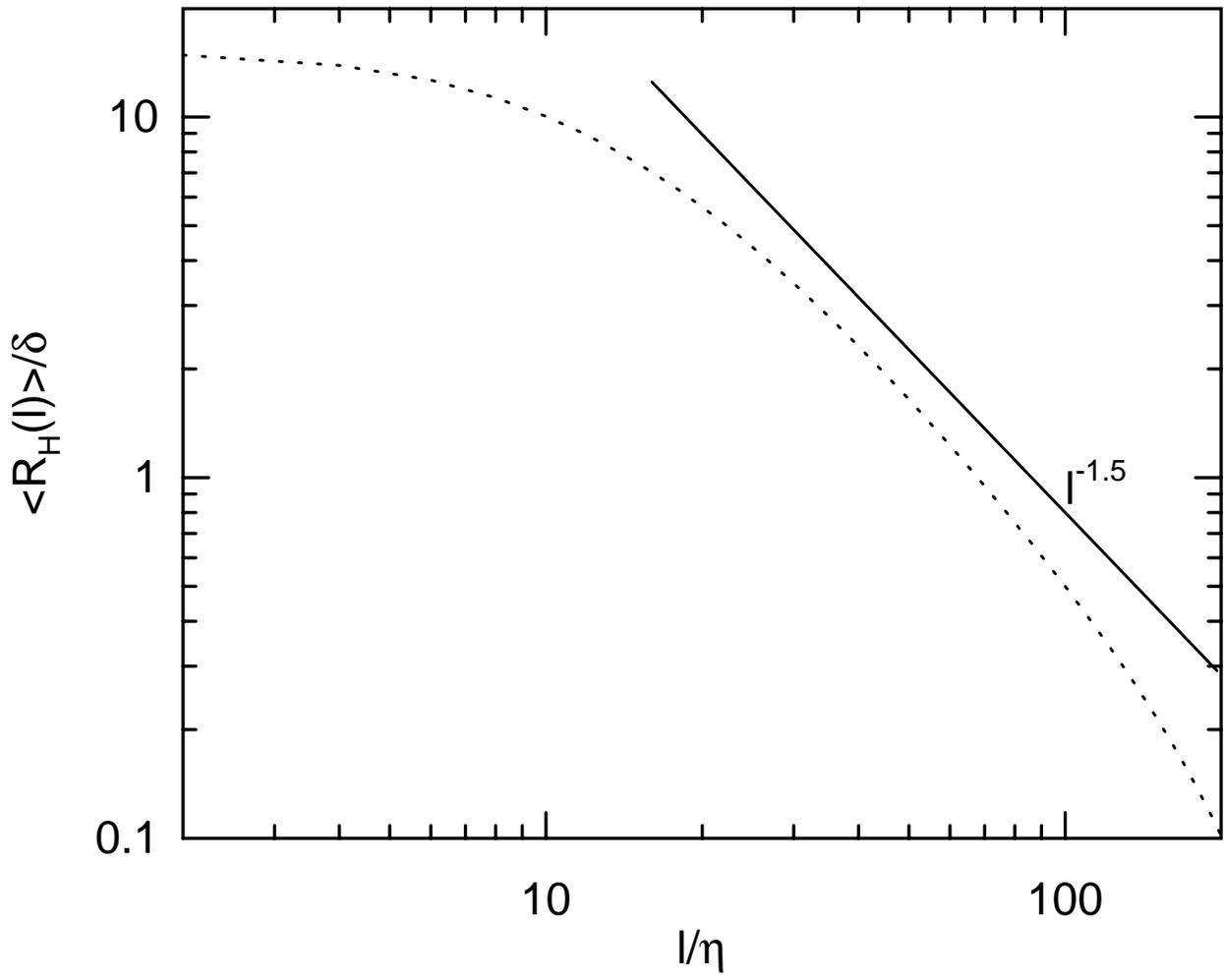
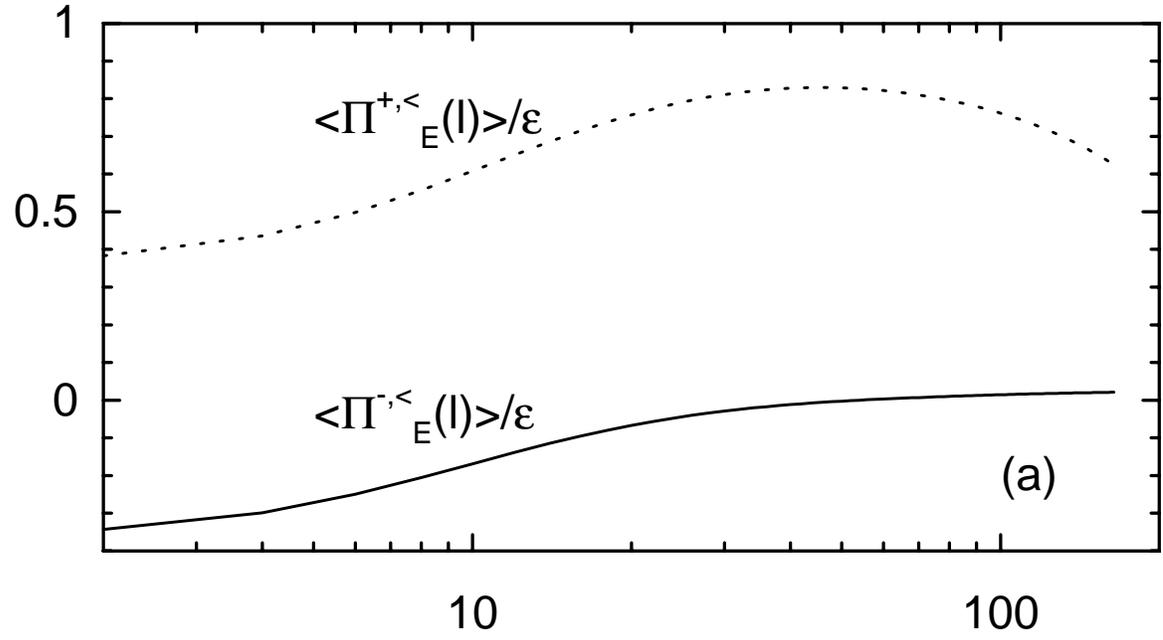
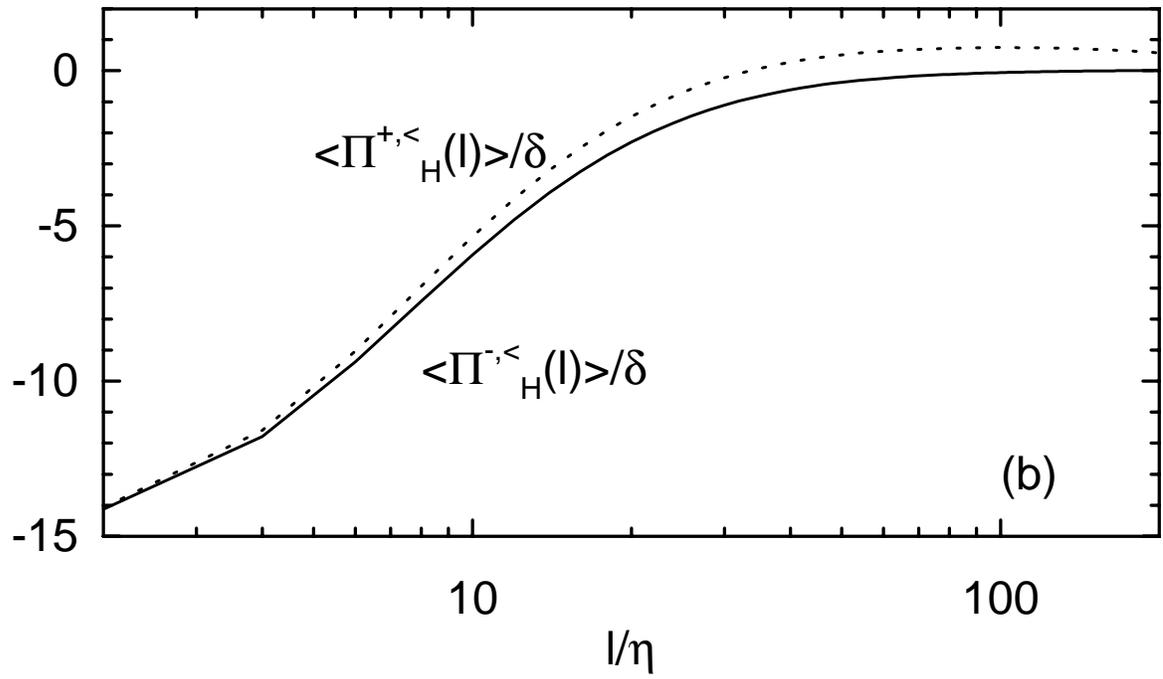
\end{document}